\documentclass[a4paper,fleqn]{cas-sc}

\usepackage[numbers]{natbib}
\usepackage[center]{caption}
\usepackage[flushleft]{threeparttable}
\usepackage{todonotes}
\usepackage[most]{tcolorbox}
\definecolor{main}{HTML}{757677}  
\definecolor{sub}{HTML}{DEE0E3}
\usepackage{framed}
\colorlet{shadecolor}{gray!15}
\colorlet{framecolor}{black}
{\endMakeFramed}

\newtcolorbox{boxH}{
    colback = sub, 
    colframe = main, 
    boxrule = 0pt, 
    leftrule = 6pt 
}

\def\tsc#1{\csdef{#1}{\textsc{\lowercase{#1}}\xspace}}
\tsc{WGM}
\tsc{QE}

\begin{document}
\let\WriteBookmarks\relax
\def\floatpagepagefraction{1}
\def\textpagefraction{.001}
\shorttitle{BugMentor: Generating Answers to Follow-up Questions from Software Bug Reports}
\shortauthors{Mukherjee and Rahman}
\title [mode = title]{BugMentor: Generating Answers to Follow-up Questions from Software Bug Reports using Structured Information Retrieval and Neural Text Generation} 

\author{Usmi Mukherjee}[orcid=0000-0001-6950-8900]
\ead{usmi.mukherjee@dal.ca}

\author{Mohammad Masudur Rahman}[orcid=0000-0003-3821-5990]
\ead{mh264771@dal.ca}

\affiliation{Dalhousie University, Halifax, Nova Scotia, Canada}

\begin{abstract}
Software bug reports often lack crucial information (e.g., steps to reproduce), which makes bug resolution challenging. Developers thus ask follow-up questions to capture additional information. However, according to existing evidence, bug reporters often face difficulties answering them, which leads to the premature closing of bug reports without any resolution. Recent studies suggest follow-up questions to support the developers, but answering the follow-up questions still remains a major challenge. In this paper, we propose BugMentor, a novel approach that combines structured information retrieval and neural text generation (e.g., Mistral) to generate appropriate answers to the follow-up questions. Our technique identifies the past relevant bug reports to a given bug report, captures contextual information, and then leverages it to generate the answers. We evaluate our generated answers against the ground truth answers using four appropriate metrics, including BLEU Score and Semantic Similarity. We achieve a BLEU Score of up to 72 and Semantic Similarity of up to 92 indicating that our technique can generate understandable and good answers to the follow-up questions according to Google’s AutoML Translation documentation. Our technique also outperforms four existing baselines with a statistically significant margin. We also conduct a developer study involving 23 participants where the answers from our technique were found to be more accurate, more precise, more concise and more useful.
\end{abstract}


\maketitle

\section{Introduction}
\looseness=-1
Software bugs are human-made errors in a software system that prevent the software from working as expected~\cite{ieeestandardglossaryforse}. Studies have shown that software bugs cost the global economy billions of dollars every year~\cite{britton2013reversible, zou2018practitioners}. Developers also spend $\sim$50\% of their programming time finding and fixing bugs~\cite{britton2013reversible}. Thus, bug resolution has been one of the major challenges in software maintenance~\cite{zou2018practitioners}. Hundreds of software bugs are submitted to bug-tracking systems like GitHub and JIRA as \textit{bug reports} ~\cite{anvik2006should}. These bugs are then triaged, analyzed, and resolved by developers.\par

\looseness=-1
Ideally, bug reports should contain all the information, such as system configuration, expected behaviour, observed behaviour, and reproducing steps that help a developer resolve a bug~\cite{chaparro2017detecting}. However, in practice, they often do not contain all the required information for reproducing or resolving the bug~\cite{chaparro2017detecting}. According to existing literature~\cite{chaparro2017detecting},  64.8\% of bug reports do not contain any expected behaviour of target software systems, and 48.6\% of them do not explicitly describe the steps to reproduce a bug. Missing information like this has been found to be a key factor behind the non-reproducibility of software bugs~\cite{rahman2020some}. In a survey conducted by Zou et al.~\cite{zou2018practitioners}, 77\%  of 327 professional developers from the major technology companies (e.g., Google, Meta, Amazon, Microsoft) consider missing information as a major problem and emphasize on complementing bug reports with useful information (e.g., steps to reproduce, environmental configuration)~\cite{zou2018practitioners}. In short, missing information has been a key challenge toward cost-effective bug resolution. \par

\looseness=-1
Many software projects on GitHub now require bug reports to adhere to specific templates or standard guidelines~\cite{githubguidelines}. Unfortunately, many bug reporters might fail to comply with them and might not be able to provide all the information during report submission~\cite {imran2021automatically}. Developers thus often pose \textit{follow-up} questions to bug reporters soliciting the missing information. Unfortunately, the bug reporters often find it challenging to answer the follow-up questions in a timely fashion, according to a recent developer survey ~\cite{rahman2020some}. For instance, Fig.\ref{fig:motivating_example} shows how a bug report was closed prematurely without any resolution due to a lack of responses to the follow-up question posed by the developer.\par

\looseness=-1
Over the last few decades, extensive research has been conducted to support various bug report management tasks, including bug triage~\cite{zhao2019unified,bodden2017proceedings}, duplicate bug detection ~\cite{nguyen2012duplicate, chaparro2019reformulating}, bug localization~\cite{zhang2019finelocator,xiao2019improving}, and bug reproduction~\cite{johnson2022empirical}. These studies have significantly advanced our understanding of the challenges in bug report management. However, important communications within bug reports, especially follow-up questions and answers between bug reporters and developers, have not been exhaustively investigated to date.  
Breu et al.~\cite{breu2010information} were the first to analyze follow-up questions from bug reports quantitatively and qualitatively. Their study revealed that 32.34\% of these questions were never responded to, which is critical for effective bug triaging, finding, and reproduction. More recently, Imran et al.~\cite{imran2021automatically} automatically recommend follow-up questions for deficient bug reports employing information retrieval. While these studies have made valuable contributions by analyzing or recommending the follow-up questions, they do not answer them, which remains a major challenge for bug reporters~\cite{rahman2020some}. In this work, we attempt to bridge this gap by automatically answering the follow-up questions where the bug reporters struggle. Our goal is to add complementary information to deficient bug reports automatically and thus support the bug report management tasks.

\looseness=-1
Automated Question Answering (QA) has been an active research topic for decades in Information Retrieval (IR) and Natural Language Processing (NLP) communities~\cite{ravichandran2002learning, brill2002analysis,waltz1978english, iyyer2014neural,asaduzzaman2013answering, tian2017apibot, lu2021beat, bansal2021neural, xu2017answerbot, abdellatif2020msrbot}. There also have been several works in the context of software engineering. Tian et al.~\cite{tian2017apibot} designed APIBot that can answer questions related to an API by analyzing relevant API documentation. Bansal et al.~\cite{bansal2021neural} designed a context-aware QA system to answer basic questions about subroutines. Lu et al.~\cite{lu2021beat} proposed a QA approach that can provide answers by executing structured queries generated from bug templates. However, their approach might fail when a bug report does not contain the requested information. Xu et al.~\cite{xu2017answerbot} designed AnswerBot that can synthesize answers for a non-factoid technical question from StackOverflow Q\&A website. While the above approaches are a source of inspiration, they do not answer the follow-up questions posed by developers against the bug reports.\par

\looseness=-1
In this paper, we propose a novel technique -- \textit{BugMentor} -- that can offer relevant answers to follow-up questions from bug reports by combining structured information retrieval and neural text generation. First, we capture textually relevant questions, answers, and bug reports against a follow-up question using structured information retrieval ~\cite{saha2013improving}. Then we capture each item's embeddings using CodeBERT~\cite{feng2020codebert} and re-rank them based on their semantic relevance to the question. Second, we leverage the ranked items above as \textit{context} with a neural text-generation technique (e.g., Mistral, Llama) and generate meaningful answers to the follow-up question~\cite{yu2022code}.\par

\looseness=-1
We select the top 20 popular projects from GitHub that use four programming languages(e.g. Java, Python, JavaScript and C++) and collect 30,869 bug reports from them for our experiments. We evaluate our technique using four popular metrics for text generation, namely BLEU score~\cite{papineni2002bleu}, METEOR~\cite{banerjee2005meteor}, Semantic Similarity~\cite{haque2022semantic}, and ROUGE~\cite{lin2004rouge}. We achieve a BLEU score of 36.18 which indicates that our generated answers are \textit{understandable to good} according to Google AutoML documentation~\cite{automldoc}. We also conduct an ablation study to justify our combination of structured information retrieval and neural text generation in BugMentor. We find that BugMentor leverages the rich context captured through structured information retrieval and thus can generate meaningful answers. BugMentor also outperforms all four baselines -- BM25~\cite{mccandless2010lucene}, Mistral~\cite{jiang2023mistral}, AnswerBot~\cite{xu2017answerbot}, DialBERT~\cite{li2020dialbert} in all four metrics.  To further demonstrate its benefit, we conduct a developer study involving 23 participants. According to the participants, the answers from BugMentor were more accurate, precise, concise and useful compared to the baseline answers.\par

\looseness=-1
To summarize, we make three contributions in this work:
\begin{itemize}
\item[(a)] A novel technique -- BugMentor -- that can generate relevant answers to follow-up questions from bug reports by combining structured information retrieval and neural text generation (e.g., Mistral).
\item[(b)] A comprehensive evaluation and validation of BugMentor using both popular performance metrics (e.g., BLEU score, METEOR score, ROUGE, Semantic Similarity) and a developer study involving 23 participants.
\item[(c)] A replication package that includes our working prototype, experimental dataset, and other configuration details for the replication or third-party reuse by the community~\cite{replicationpackage}.
\end{itemize}

\section{Motivating Example}

\begin{figure}[!ht]
  \centering
\includegraphics[width=\textwidth,height=4.5in, keepaspectratio]{motivating_example_1.pdf}
  \caption{An example of a bug report (ID \#58280) being closed due to a lack of response to the follow-up question}
  \label{fig:motivating_example}
\end{figure}

\looseness=-1
To demonstrate the potential benefit of our work, let us consider the example bug report in Fig.~\ref{fig:motivating_example}. It has been taken from the \emph{tensorflow} repository on GitHub~\cite{MotivatingExampleTensorflow}. The bug report (Step (a), Fig~\ref{fig:motivating_example}) discusses a memory allocation issue and provides minimal steps to reproduce the issue. However, it does not follow the standard issue template suggested by GitHub~\cite{issueTemplate_Github} and lacks sufficient information about the system configuration. Subsequently, the developer (\textit{@mohantym}) requests the missing information as a follow-up question that focuses on the occurrence of the issue on the \textit{nightly} version of the software (Step (b), Fig.~\ref{fig:motivating_example}). However, neither the bug reporter (\textit{@DanielZanchi}) nor anyone else provided an answer to the follow-up question. As a result, the bug report was initially marked as \textit{stale} (Step (c), Fig.~\ref{fig:motivating_example}) and later closed due to a lack of response (Step (d), Fig.~\ref{fig:motivating_example}).\par
As shown in Fig~\ref{fig:motivating_example}, without proper support in gathering missing information, software bugs either take longer to be resolved or remain unresolved and are ultimately closed. Several bugs, such as memory allocation bugs, are considered to be severe~\cite{li2006bugseverity} that significantly impact software quality. Our work --- BugMentor --- delivers meaningful answers to such follow-up questions. \par

\looseness=-1
To illustrate the effectiveness of BugMentor, let us examine the bug report Fig.~\ref{fig:motivating_example2} submitted to the \textit{TypeScript} repository on GitHub~\cite{MotivatingExampleTypeScript}. The bug report discusses an issue where TypeScript compiler fails to throw an error when encountering generic type parameters in JavaScript code. TypeScript compiler can compile both JavaScript and TypeScript code. However, generic type parameters are TypeScript features and are not allowed in JavaScript code. The expected behavior for TypeScript is to issue an error in this context, as generic type parameters should not be allowed in JavaScript code. However, the actual behavior is that no error is thrown (Step (a), Fig~\ref{fig:motivating_example2}). The developer asks a follow-up question inquiring about TypeScript's handling of the type parameter in this context, whether it simply ignores the type parameter or if it contributes to type inference and checking in some way(Step (b), Fig~\ref{fig:motivating_example2}).\par

\renewcommand{\arraystretch}{1.3}
\begin{table}[]
\centering
\caption{Bug Report Explanation}
\label{bugmentor-ans}
\begin{tabular}{|p{0.9\columnwidth}|} 
\hline
\rowcolor{gray!15} 
\multicolumn{1}{|c|}{\textbf{\textcolor{black}{Bug Report~\hyperref[bugmentor-ans]{\#31683}}}} \\ \hline \hline
\textbf{(a) Answer:} TypeScript should validate constructs in JS files but throw an error for TS-specific features like type parameters. In this case, the type parameter is not ignored; instead, a new type variable is created within the static method's scope, shadowing the generic type parameter. This may cause unexpected type inference and checking results. \\ \hline
\textbf{(b) Accepted Answer:} For clarity, we parse and check most TypeScript constructs even in JavaScript files. However, we are supposed to issue an error stating that "X is not valid in a JavaScript file," where in this case, X refers to type parameters. \\ \hline
\end{tabular}
\end{table}

BugMentor's response provides valuable insights into the issue at hand in Table~\ref{bugmentor-ans}. 
It acknowledges that while TypeScript can be used to validate JavaScript code, it currently fails to issue errors when encountering invalid features such as type parameters in JavaScript code. Instead, it incorrectly allows the creation of a new type variable that shadows the original, potentially leading to type inference errors. The response generated by BugMentor highlights the need for better error reporting to prevent the misuse of TypeScript features in JavaScript code. This aligns with the accepted answer, which confirms that TypeScript should issue an error stating that ``X is not valid in a JavaScript file," where ``X" refers to type parameters in this case. Thus, BugMentor's answer not only provides additional context by explaining how TypeScript currently handles the type parameters but also helps the developer understand the root cause of the issue. With this information, the developer can now focus on reproducing and fixing the issue (e.g., misuse of TypeScript features in JavaScript code).

\begin{figure}[!t]
    \centering
    \includegraphics[width=\textwidth, height=5.5in,keepaspectratio]{motivating-example-2.pdf}
    \caption{An example bug report(ID \#31683) with a follow-up question }
    \label{fig:motivating_example2}
\end{figure}


\section{BugMentor : Proposed Technique}

\begin{figure}
    \centering
    \includegraphics[width=0.8\linewidth]{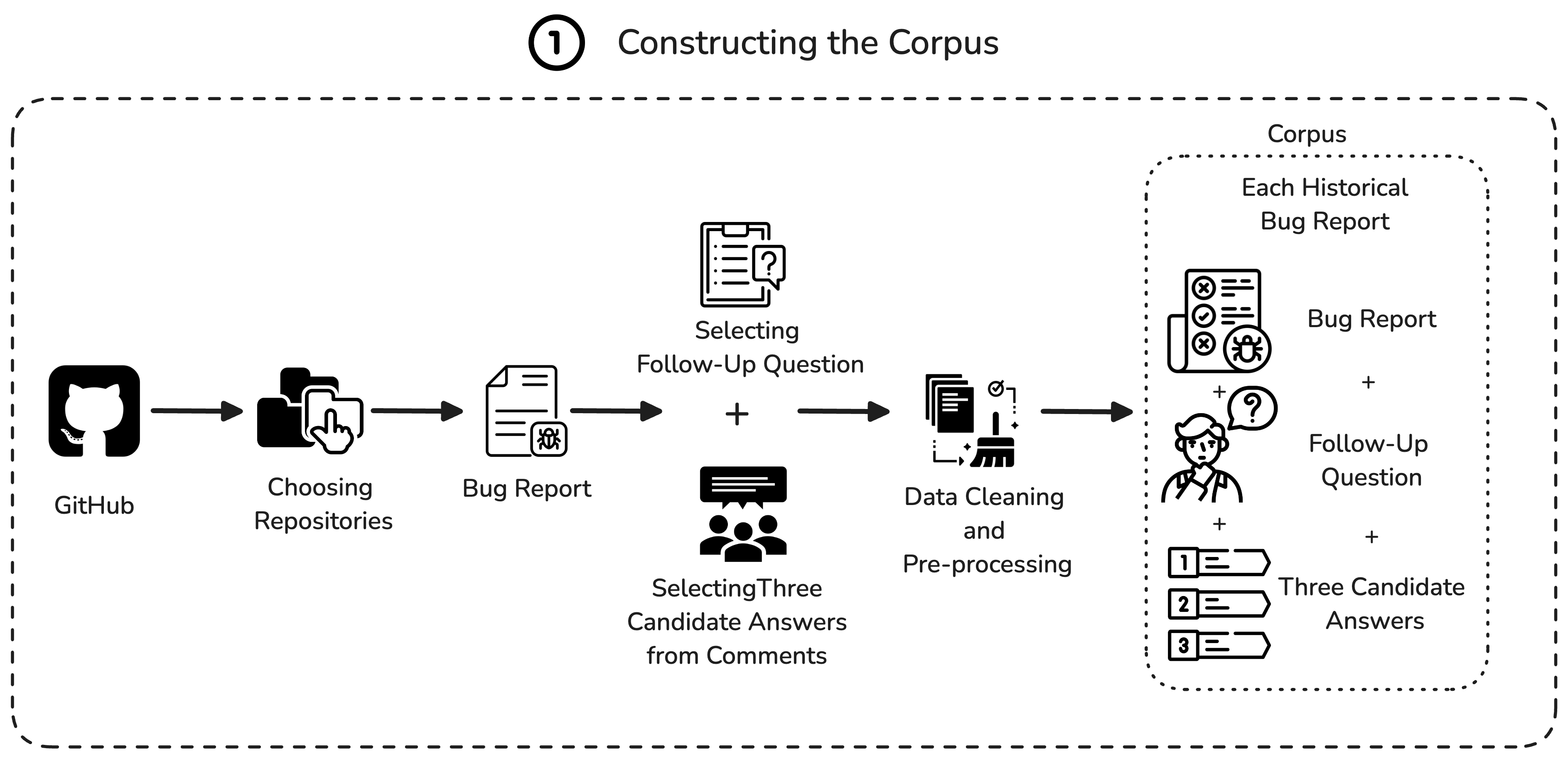}
    \caption{Constructing the Corpus}
    \label{fig:algo-step1}
\end{figure}

\looseness=-1
Fig. ~\ref{fig:algo-step1}, ~\ref{fig:algo-step2} and ~\ref{fig:algo-step3-4} show Steps 1, 2, 3 and 4 of the schematic diagram of our proposed technique --- BugMentor. As input, it accepts a bug report of interest, its follow-up question, and a corpus of past bug reports with their follow-up questions and corresponding answers. As the output, our technique generates a relevant answer to the follow-up question. We discuss different steps of our technique in detail in the following sections.\par

\subsection{Constructing the Corpus}
\label{sec:corpus-construction}
\looseness=-1
We construct our corpus using past bug reports and their discussion history from 20 real-world, open-source software systems on GitHub (Step 1, Fig~\ref{fig:algo-step1}).\par

\subsubsection{\textit{Choosing the Repositories for the Corpus}}
\label{sec:proposedtechnique-repo}
To construct our corpus, we collect high-quality repositories using a semi-automated approach. We follow the approach of Imran et al. ~\cite{imran2021automatically} and choose GitHub~\cite{github} as a data source. GitHub~\cite{github} is a popular open-source platform that supports various software maintenance practices, including bug report management. We select our repositories from GitHub as follows.\par

First, we select the most starred, active repositories containing a minimum of 500 issues (reported as of May 2023) using GitHub's advanced search~\cite{GitHubAdvancedSearch}. The most active repositories with the most number of issues belonged to four programming languages --- Python, Java, Javascript and C++. We then categorize the repositories into four subsets based on their programming languages, where each programming language had five repositories. The repositories that belonged to other programming languages that didn't have atleast five repositories with the minimum number of issues were discarded.\par
\begin{figure}
    \centering
    \includegraphics[width=\textwidth]{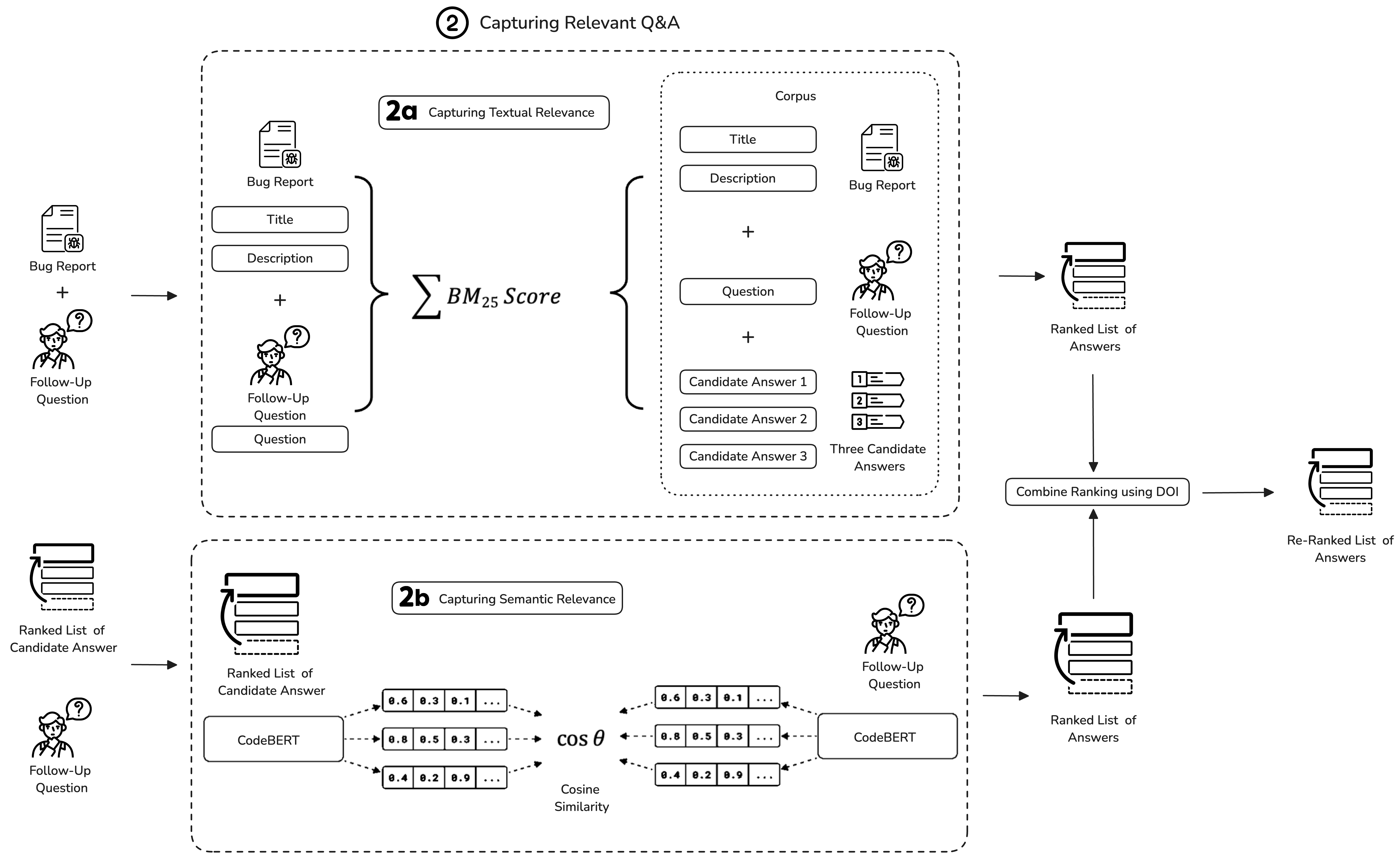}
    \caption{Capturing Relevant Answers}
    \label{fig:algo-step2}
\end{figure}
\subsubsection{\textit{Choosing the Bug Reports for the Corpus}}
\looseness=-1
From each repository, we select the issues that were closed within the last five years to ensure the issues were recent and there hasn't been a concept drift. We also select the issues that are labelled as ``bug", ``crash", or ``defect" to ensure that they are discussing software bugs or defects. We also select the bug reports labelled as ``needs more info" and ``stale", indicating that they were closed due to a lack of activity. We use GitHub's REST API~\cite{githubdocumentation} to collect the bug reports and their discussion history. Each of our collected bug reports consists of several fields, namely issue ID, title, bug description, bug reporter, label, creation time, and resolution time. \par

\subsubsection{\textit{Selecting Follow-up Questions}}
\label{sec:proposedtechnique-followq}
To select follow-up questions from each bug report, we first collect their issue comments using a GitHub API client~\cite{koshukeGitHubLibrary}. From each comment, we capture four different fields, namely comment ID, author of the comment, comment, and comment time. Following the strategy of Imran et al. ~\cite{imran2021automatically}, we collect the comments that begin with an interrogative word and end with a question mark. We use NLTK's Classifier~\cite{nltkclassifydocumentation} to identify these comments, as was applied previously~\cite{yuan2012enhancing}. We also consider comments that requested additional information using words such as ‘please’ or `can you' as valid comments. Then we select the first valid, interrogative comment that is not written by the bug reporter as our follow-up question from each bug report. We manually check up to 30 comments from each bug report to identify our follow-up questions. \par

\subsubsection{\textit{Selecting Candidate Answers}}
\label{sec:proposedtechnique-candidateans}
The next step in our corpus construction is to select candidate answers against the follow-up questions above. 
We apply three criteria to the selection of our candidate answers: (a) Candidate Answer 1 --- the first comment after the follow-up question that was not authored by the question submitter ~\cite{imran2021automatically}, (b) Candidate Answer 2 --- the first comment after the follow-up question that was authored by the bug reporter, and (c) Candidate Answer 3 --- the most similar comment to the follow-up question based on the BM25 algorithm~\cite{whissell2011improving} as the comment addressing the question would be textually similar to the question. Finally, our corpus consisted of hundreds of bug reports where we capture the Bug ID, title, description, follow-up question and three candidate answers from each bug report.\par

\subsubsection{\textit{Data Pre-processing}}
\label{sec:proposedtechnique-datapreprocessing}
\looseness=-1
We apply standard natural language pre-processing to each bug report, follow-up question and candidate answer from our corpus. We remove redundant or noisy elements such as escape sequences, special characters, stack traces or images from each item~\cite{imran2021automatically}  to facilitate better analysis. We retain all the stopwords and do not perform stemming to preserve the semantics of all the textual elements.\par


\subsection{Capturing Relevant Answers}
\label{sec:proposedtechnique-relanswer}
Once the corpus is constructed, we then capture relevant candidate answers against each follow-up question (Step 2, Fig~\ref{fig:algo-step2}) and its respective project-specific corpus. We use the ElasticSearch~\footnote{\url{https://www.elastic.co/elasticsearch/}}  implementation of Lucene, a widely adopted search engine combining Boolean search and Vector Space Model (VSM)~\cite{haiduc2013automatic,moreno2015query, mccandless2010lucene}, for our task. We employ the Okapi BM25 algorithm~\cite{kamphuis2020bm25} from the engine for similarity calculation. In particular, we calculate two BM25-based relevance scores where we adapt an existing work of Saha et al.~\cite{saha2013improving}:
\vspace{-0.21cm}
\begin{equation}\label{bluir}
\vspace{-0.25cm}
  s'\left( \vec{d},\vec{q} \right) = \sum_{r\in Q}\sum_{f\in D}s(d_{f} , q_{r})
\end{equation}
Here $q_{r}$ is a query representation, and $d_{f}$ is a field from the past bug report (e.g., title, description).\par

\begin{figure}
\centering
\includegraphics[width=\textwidth,height=3.5in, keepaspectratio]{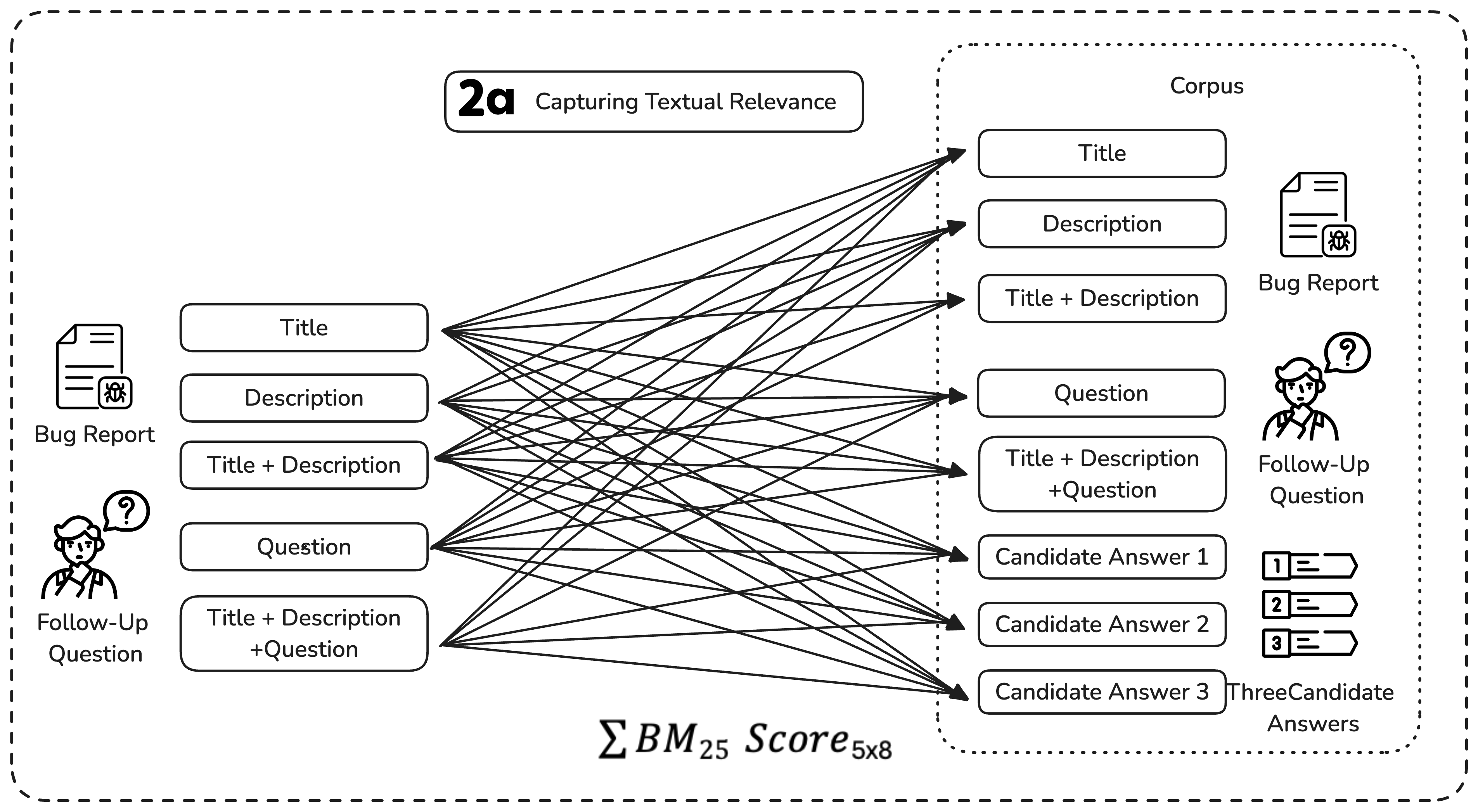}
	\caption{Capturing relevant answers with structured Information Retrieval}
	\label{fig:qa}
\end{figure}
\subsubsection{Detecting Relevant Answers}
\looseness=-1
First, we capture five items from each of the given and past bug reports --- title, description, follow-up question, title + description, and title + description + question. Second, we capture five items from the given bug report --- title, description, and follow-up question, title + description, and title + description + question and three candidate answers from each past bug report. Then, we conduct 5x8 similarity calculations using Eq.~\ref{bluir}, for each of the items, as shown in Fig.~\ref{fig:qa}. This score indicates the relevance between the follow-up question (and its bug report) and each candidate answer from the past bug report.
According to existing literature~\cite{saha2013improving}, such an element-based similarity calculation can help prevent spurious matching. This score indicates the general relevance between a given and past bug reports. This score indicates the relevance between the follow-up question (and its bug report) and each candidate answer from the past bug report. The score from Step 2a indicates a more granular relevance between the question and answer. Thus, the above scores capture the relevance of each answer against the given follow-up question (and corresponding bug report).\par

\subsubsection{Ranking Based on Textual Relevance}
We rank the candidate answers based on their BM25-based relevance scores calculated above (Step 2, Fig.~\ref{fig:algo-step2}). In particular, we capture the top K (e.g., 5) relevant candidate answers from the corpus against a follow-up question (Step 2, Fig.~\ref{fig:algo-step2}). It should be noted that these answers can come from various bug reports.\par

\subsubsection{Ranking Based on Semantic Relevance}
\looseness=-1
BM25 algorithm relies on keyword matching for relevance estimation, which could suffer from the vocabulary mismatch problem~\cite{furnas1987vocabulary}. We thus incorporate embedding-based similarity into our approach and detect the semantically relevant candidate answers. We capture word embeddings trained by CodeBERT~\cite{feng2020codebert} trained on CodeSearchNet~\cite{husain2019codesearchnet}. Candidate answers can contain both code and natural language elements. Since CodeBERT was fine-tuned for code and natural language tasks~\cite{feng2020codebert}, it can represent our candidate answers better. We calculate the cosine similarity between the word embeddings of the follow-up question and that of each candidate answer. We then re-rank the answers based on their semantic relevance to the question and return the top K answers (Step 2, Fig.~\ref{fig:algo-step2}).\par

\subsubsection{Capturing the Top Relevant Answers}
\looseness=-1
We combine both BM25-based ranking and semantic relevance-based ranking using the Degree of Interest (DOI) method. Rahman et al.~\cite{rahman2016rack} use the following formulae to combine two orthogonal rankings similar to both our rankings:
\begin{equation}
DOI = \frac{I}{N} 
\end{equation}

\looseness=-1
Here, $I$ is the position of an answer in the ranked list and $N$ is the total number of answers. First, we calculate the DOI score of each answer within the BM25-based list and then, we calculate the DOI score within the semantic relevance-based ranked list. Then we combined the DOI scores for each answer and rank the answers based on their combined DOI score using simple summation.\par


\subsection{Answering the Follow-up Question}
Once the relevant candidate answers are captured, we then construct the context and generate answers to the follow-up question using neural text generation (e.g Mistral).

\subsubsection{Leveraging Mistral for Context-Aware Answer Generation}
To generate relevant answers to follow-up questions, we employ Mistral, a state-of-the-art pre-trained decoder-only Transformer model for neural text generation. Mistral has been pre-trained on internet-scale data and adopts an encoder-decoder network to generate texts~\cite{jiang2023mistral}. We chose Mistral for our task due to its strong performance in context-aware text generation and its ability to handle long-range dependencies, which is crucial for processing bug report information~\cite{jiang2023mistral}. From a performance-to-efficiency standpoint, Mistral offers a strong performance-to-efficiency ratio while remaining relatively lightweight compared to other modern language models, maintaining competitive performance despite its smaller size~\cite{labrak2024biomistral}. In terms of technical capabilities, Mistral's advanced attention mechanisms excel at processing multiple textual components like bug reports, descriptions, follow-up questions, and candidate answers, while supporting context-aware text generation~\cite{jiang2023mistral}. Finally, from a deployment perspective, the model provides significant advantages through local deployment options without external API dependencies, flexibility for future fine-tuning, and efficient inference suitable for bug report analysis tasks~\cite{wilkins2024higher}. While our technique remains model-agnostic and can incorporate other language models as needed, these combined factors made Mistral particularly suitable for generating accurate and contextually appropriate answers to follow-up questions in bug reports.\par

\begin{figure}
    \centering
    \includegraphics[width=0.6\linewidth]{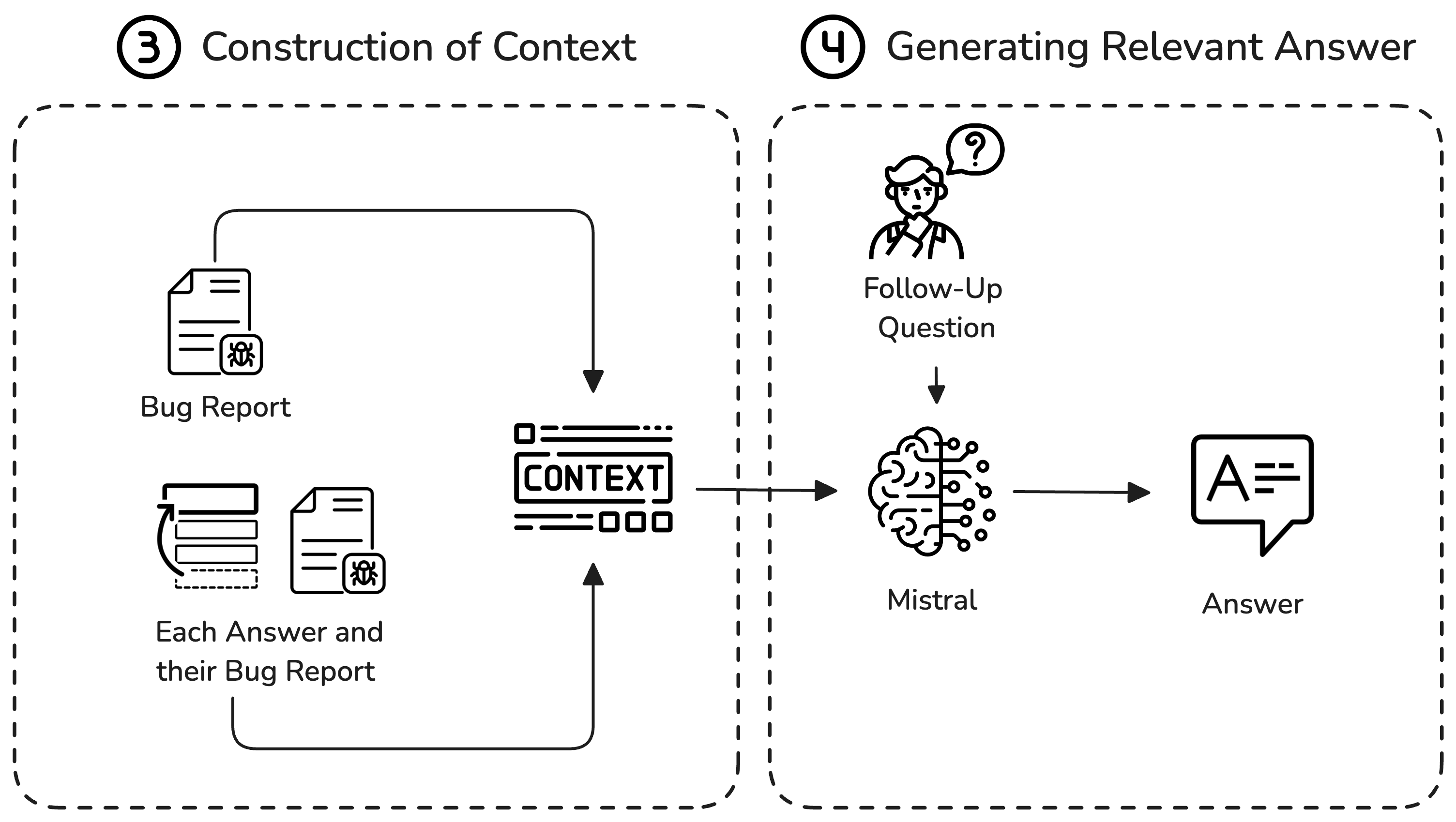}
    \caption{Answering the Follow-Up Question}
    \label{fig:algo-step3-4}
\end{figure}

\subsubsection{Constructing the context}
\looseness=-1
Context plays a vital role in enabling generative question-answering models like Mistral to comprehensively understand the semantics and intent of a question, leading to more accurate and relevant answers~\cite{huggingfacegenerativeqa,seonwoo2020context}. We then construct the context to enrich our follow-up question (Step 3, Fig~\ref{fig:algo-step3-4}) using the relevant answers from the previous step. 
Thus, to construct the context, we use three items from the previous step --- one answer from the ranked list, its bug report, and the given bug report. The answer and its bug report are likely to contain additional information to compensate for the missing information in the given bug report that triggers a follow-up question. The answer and its associated bug report likely contain additional details that can help address the follow-up question more comprehensively. We repeat the context construction for each of the K candidate answers and send them to our Mistral model for final answers.\par 

\subsubsection{Generating relevant answers }
\looseness=-1
We then generate relevant answers to the follow-up question by leveraging our context above with Mistral(Step 4, Fig~\ref{fig:algo-step3-4}). The model requires two components to operate, a question and its context. We provide the model with a follow-up question and its context from the previous step and capture an answer to the follow-up question from the model. We construct our prompt for our generation task following existing literature~\cite{lazaridou2022internet, white2023prompt} using chain of thought prompting~\cite{replicationpackage}. We designed a structured prompt to guide the language model in answering follow-up questions about bug reports. The prompt template is shown in Table~\ref{prompt-template}. For example, our technique --- BugMentor --- offers the answer against the example question in Fig.~\ref{fig:motivating_example} in Table~\ref{bugmentor-ans}.\par


\renewcommand{\arraystretch}{1.3}
\begin{table}[!ht]
\centering
\caption{Prompt Template}
\label{prompt-template}
\begin{tabular}{|p{0.9\columnwidth}|} 
\hline
\rowcolor{gray!15} 
\multicolumn{1}{|c|}{\textbf{Prompt Template}} \\ \hline \hline
Answer Questions on the bug report based on the relevant information. \\
Here is a bug report which has incomplete information. \\
\#\#Bug Report - \\
\{bug\_report\} \\

There is a follow-up question asking for missing information. \\
Here is some relevant information from a previous bug report. \\
\{relevant\_candidate\_answer\} \\

Can you answer the question below based on the bug report? \\
Here is the question. \\

\#\# Question- \\
\{question\} \\

\#\# Answer : \\
\hline
\end{tabular}
\end{table}


\section{Study Methodology}
\looseness=-1
We curated a large dataset containing 30,869 bug reports and their follow-up questions from 20 subject systems and evaluate BugMentor using four appropriate metrics from relevant literature --- BLEU~\cite{papineni2002bleu}, METEOR ~\cite{banerjee2005meteor}, Semantic Similarity~\cite{haque2022semantic} and ROUGE~\cite{lin2004rouge}. To place our work in the literature, we also conduct an ablation study~\cite{mccandless2010lucene,jiang2023mistral} and compare our technique with three baseline techniques. 
Through our experiments, we answer four research questions as follows:
\begin{enumerate}
\item[(a)] \textbf{RQ$\mathbf{_1}$}: How does BugMentor perform in answering follow-up questions in terms of different automatic evaluation metrics?
\item[(b)] \textbf{RQ$\mathbf{_2}$}: Can BugMentor outperform the existing baselines in terms of automatic evaluation metrics?
\item[(c)]\textbf{RQ$\mathbf{_3}$}: How do different components impact the overall performance of BugMentor?
\item[(d)]\textbf{RQ$\mathbf{_4}$}: How does BugMentor perform in answering follow-up questions when compared with ChatGPT?
\item[(e)] \textbf{RQ$\mathbf{_5}$}: How accurate, precise, useful, and concise are the answers from BugMentor? 
\end{enumerate}

\subsection{Dataset Construction}\label{sec:dataset-cons}

We first construct our corpus and ground truth for conducting the experiments.

\subsubsection{Corpus Construction}

\looseness=-1
To construct our corpus, we chose the top 20 popular projects from GitHub written in 4 different programming languages and collected 30,869 bug reports from them. We also capture a follow-up question and three candidate answers from each bug report. Finally, our corpus consisted of hundreds of bug reports where we capture the Bug ID, title, description, follow-up question and three candidate answers from each bug report. We ensure that the candidate answers are selected from bug reports in the corpus submitted prior to the given bug report. We apply standard natural language pre-processing to each item from our corpus. Please check Section~\ref{sec:corpus-construction} for further details on corpus construction.\par

\subsubsection{Ground-Truth Construction}
\label{sec:groundtruth}

\looseness=-1
To evaluate BugMentor, we first construct a randomly sampled, held-out dataset (i.e., 95\% confidence level and 4.06\% error margin) containing 550 bug reports ($\sim$27 bug reports x 20 systems). We then involve six human annotators (e.g., graduate students) to determine the ground truth answers against the follow-up questions from each bug report. The human annotators each have 2-3 years of experience, including 1-2 years of industry experience in the four programming languages used in our scenario. We divided 550 bug reports into six buckets (Table~\ref{tab:interannotatoragreement}), each containing $\sim$90 bug reports, their questions, and candidate answers. Each bucket was annotated by three annotators, resulting in $\sim$270-275 bug reports per annotator. Then, we used majority voting ~\cite{kuhrmann2017pragmatic} to determine the ground truth answers. That is, the answer having the majority of votes was chosen as the ground truth answer against a follow-up question. When the answers did not have a clear majority, i.e. for 3\% of the dataset, the three annotators engaged in discussions to resolve conflicts and determine the ground truth answer~\cite{kuhrmann2017pragmatic}. Each annotator spent $\sim$2.5-3 hours to complete the annotation task. \par

\looseness=-1
We compute the Cohen's \emph{$\kappa$} for all pairs of annotators, and the result is reported in Table~\ref{tab:interannotatoragreement}. Although we use majority voting for annotation, our calculated metrics show the agreement level for each pair of annotators. We found an average of 0.46, which indicates a moderate agreement between any two annotators.\par

\begin{table}[!h]
\centering
\caption{Inter-annotator Agreement}
\label{tab:interannotatoragreement}
\resizebox{\textwidth}{!}{
\begin{tabular}{|cc|cc|cc|cc|cc|cc|}
\hline
\multicolumn{2}{|c|}{\textbf{Bucket 1}}     & \multicolumn{2}{c|}{\textbf{Bucket 2}}     & \multicolumn{2}{c|}{\textbf{Bucket 3}}     & \multicolumn{2}{c|}{\textbf{Bucket 4}}     & \multicolumn{2}{c|}{\textbf{Bucket 5}}     & \multicolumn{2}{c|}{\textbf{Bucket 6}}     \\ \hline
\multicolumn{1}{|c|}{\textbf{Pairs}} & \textbf{$\kappa$}    & \multicolumn{1}{c|}{\textbf{Pairs}} & \textbf{$\kappa$}    & \multicolumn{1}{c|}{\textbf{Pairs}} & \textbf{$\kappa$}    & \multicolumn{1}{c|}{\textbf{Pairs}} & \textbf{$\kappa$}    & \multicolumn{1}{c|}{\textbf{Pairs}} & \textbf{$\kappa$}    & \multicolumn{1}{c|}{\textbf{Pairs}} & \textbf{$\kappa$}    \\ \hline \hline
\multicolumn{1}{|c|}{A1 \& A2}       & 0.72 & \multicolumn{1}{c|}{A2 \& A3}       & 0.58 & \multicolumn{1}{c|}{A3 \& A4}       & 0.44 & \multicolumn{1}{c|}{A4 \& A5}       & 0.66 & \multicolumn{1}{c|}{A5 \& A6}       & 0.49 & \multicolumn{1}{c|}{A1 \& A2}       & 0.41 \\ \hline
\multicolumn{1}{|c|}{A2 \& A3}       & 0.24 & \multicolumn{1}{c|}{A3 \& A4}       & 0.43 & \multicolumn{1}{c|}{A4 \& A5}       & 0.53 & \multicolumn{1}{c|}{A5 \& A6}       & 0.40 & \multicolumn{1}{c|}{A1 \& A5}       & 0.44 & \multicolumn{1}{c|}{A2 \& A6}       & 0.29 \\ \hline
\multicolumn{1}{|c|}{A3 \& A1}       & 0.39 & \multicolumn{1}{c|}{A4 \& A2}       & 0.58 & \multicolumn{1}{c|}{A5 \& A3}       & 0.65 & \multicolumn{1}{c|}{A6 \& A4}       & 0.27 & \multicolumn{1}{c|}{A6 \& A1}       & 0.49 & \multicolumn{1}{c|}{A6 \& A1}       & 0.40 \\ \hline
\end{tabular}}
\end{table}

\subsection{Evaluation Metrics}

\looseness=-1
To evaluate BugMentor's answers against the ground truth, we use four relevant metrics from literature --- BLEU Score~\cite{papineni2002bleu}, METEOR Score~\cite{banerjee2005meteor}, ROUGE Score~\cite{lin2004rouge} and Semantic Similarity metric~\cite{haque2022semantic}. They are defined as follows:\par

\subsubsection{\emph{\textit{ BLEU --- Bi-Lingual Evaluation of Understudy: }}}
\looseness=-1
BLEU score is a frequently used metric for evaluating translation~\cite{papineni2002bleu}, which has found application in many software engineering tasks (e.g., comment generation~\cite {hu2020deep}, automatically explaining the bugs~\cite{mahbub2023explaining}). It compares a candidate text to a reference text and determines how similar they are based on the matching of their n-grams. The BLEU score is calculated as follows:\par
\begin{equation}
BLEU = BP \cdot exp \left ( \sum_{n=1}^{N}w_{n}log(p_{n}) \right )
\end{equation}

Here $N$ is the maximum n-gram order, $w\textsubscript{n}$ is the weight assigned to the n-gram order, and $BP$ is the brevity penalty - a factor that penalizes the BLEU score when the candidate text is shorter than the reference text. $p\textsubscript{n}$ is the modified n-gram precision, which measures the ratio of the overlapping n-grams (between the candidate text and the reference text), and the total number of n-grams in the candidate text.\par

\subsubsection{\emph{ \textit{ METEOR --- Metric for Evaluation of Translation with Explicit ORdering: }}}
\looseness=-1
The METEOR score is a metric for evaluating the quality of machine translation output based on both lexical and syntactic information~\cite{banerjee2005meteor}. It measures the similarity between a candidate text and the reference text by sequentially applying exact match, stemmed match and wordnet-based synonym match between the texts. The METEOR score is computed as a combination of precision and recall of the matches, with a higher emphasis on recall. The formula for METEOR is given by:
\begin{equation}
METEOR = F_{mean} \cdot (1 - Penalty)
\end{equation}
where $F_{mean}$ is the harmonic mean of precision $P$ and recall $R$, defined as:
\begin{equation}
F_{mean} = \frac{10 \cdot P \cdot R}{R + 9 \cdot P}
\end{equation}
The penalty is a function of the chunkiness of the matched segments and is calculated as:
\begin{equation}
Penalty = 0.5 \cdot \left(\frac{\# \text{chunks}}{\# \text{unigrams in matches}}\right)^3
\end{equation}

This ensures that translations with fewer chunks of contiguous matched words are favored, reflecting higher order of correctness in terms of both semantics and syntax.

\subsubsection{\emph{\textit{ ROUGE --- Recall-Oriented Understudy for Gisting Evaluation}}}

\looseness=-1
ROUGE is a set of metrics designed to evaluate generated text to one or more reference texts~\cite{lin2004rouge}. It measures the overlap of n-grams, word sequences, and word pairs between the candidate and reference texts. The most commonly used variants are ROUGE-N, ROUGE-L, and ROUGE-W. For our task, we use ROUGE-L, which measures the longest common subsequence (LCS) between the candidate and reference texts. ROUGE-L is more flexible as it doesn't require consecutive matches and can capture sentence-level word order effectively. The ROUGE-L score is calculated as follows:

\begin{equation}
R_{lcs} = \frac{LCS(X,Y)}{m}, \quad P_{lcs} = \frac{LCS(X,Y)}{n}
\end{equation}

\begin{equation}
F_{lcs} = \frac{(1 + \beta^2)R_{lcs}P_{lcs}}{R_{lcs} + \beta^2P_{lcs}}
\end{equation}

Where $X$ is the reference summary of length $m$, $Y$ is the candidate summary of length $n$, $LCS(X,Y)$ is the length of the longest common subsequence between $X$ and $Y$, and $\beta$ is usually set to favor recall ($\beta > 1$). ROUGE-L provides a comprehensive evaluation of summary quality by capturing both content overlap and sequence similarity, making it particularly suitable for our task of assessing generated text against reference texts.

\subsubsection{\emph{\textit{ SS --- Semantic Similarity: }}}

\looseness=-1
In a recent work, Haque et al.~\cite{haque2022semantic} investigate which metric reflects human assessment of similarity the best. They use Sentence-BERT~\cite{reimers2019sentence} to capture meaningful sentence embeddings. Then they compare when a candidate text with the reference text based on these embeddings using cosine-similarity. According to Haque et al. such a similarity, it has the highest correlation with human-evaluated similarity. Semantic similarity can be computed as follows:

\begin{equation}
SemSim(ref, gen) = \cos(\text{sbert}(ref), \text{sbert}(gen))  
\end{equation}

where $sbert(ref)$, $sbert(gen)$ are the numerical representations from Sentence-BERT for the reference text and generated text respectively.\par

\subsection{Baseline Selection}
To the best of our knowledge, there exists no work that can offer relevant answers to follow-up questions from bug reports. However, there have been traditional methods and tools to support Q\&A in several other contexts. Lucene~\cite{mccandless2010lucene} is a popular IR-based tool that has been used to recommend answers in the programming Q\&A site ~\cite{Atwood_2017,ponzanelli2013leveraging}.
Mistral~\cite{jiang2023mistral} is a large language model for generating answers to general-purpose questions. AnswerBot~\cite{xu2017answerbot} can synthesize answers for technical, non-factoid questions on StackOverflow. 
DialBERT~\cite{li2020dialbert} can analyze mixed conversations using BERT and BiLSTM models and find comments that are most related to a given comment. Since our task also involves finding the most relevant answer to a given question, DialBERT could be a suitable baseline for our experiments.
We thus consider these four as traditional method as baselines for our comparison. We call them Baseline$_{Lucene}$, Baseline$_{Mistral}$, AnswerBot, and DialBERT, respectively in this experiment. \par

\looseness=-1
To implement Baseline$_{Lucene}$, we provide a follow-up question as the \emph{query} and all candidate answers as the \emph{corpus} to the Lucene tool. Then we collect the top K answers from the tool by executing the query for our evaluation. For Baseline$_{Mistral}$, we provide a follow-up question as the \emph{query} and its corresponding bug report as \emph{context} to the Mistral model, which generates an answer. To implement DialBERT, we use the replication package provided by the authors~\cite{li2020dialbert}. We provide a follow-up question as the \emph{query} and all bug reports along with their candidate answers as the \emph{context}. To implement AnswerBot, we use the replication package provided by the authors~\cite{xu2017answerbot, maxxbw54}. We provide a follow-up question as the \emph{query} and all bug reports along with their candidate answers as the \emph{corpus}.\par

\section{Evaluation of BugMentor}
\subsection{Answering RQ$\mathbf{_1}$ --- How does BugMentor perform in answering follow-up questions in terms of different automatic evaluation metrics?}

In this experiment, we analyze the performance of BugMentor using four different evaluation metrics - BLEU score~\cite{papineni2002bleu}, Semantic Similarity ~\cite{haque2022semantic}, METEOR~\cite{banerjee2005meteor} and ROUGE~\cite{lin2004rouge}. We divide our held-out dataset into four subsets based on their corresponding programming languages and report the results for each subset. Table~\ref{tab:bugmentorcontext2} shows the performance details of BugMentor. It should be noted that a higher value for BLEU, METEOR, Semantic Similarity, and ROUGE metrics is desirable in our experiments.\par

\looseness=-1
BugMentor achieves an average BLEU Score of 30.48 for Top 1 answer, and a maximum of 33.15 for Top 5 answers when our whole dataset is considered. However, our technique achieves a maximum of 34.38 for the Top 1, 34.99 for Top 3 and 36.18 for Top 5 answers across all four subsets. These BLEU scores indicate that our generated answers are \textit{understandable} to \textit{good} according to Google’s AutoML Translation documentation~\cite{automldoc}. They also show that the answers generated by BugMentor have a significant overlap with the ground truth in terms of their words and word order. However, BLEU score primarily focuses on determining the precision of an answer against the ground truth answer. Hence, we also evaluate our answers using the METEOR score, where recall is calculated by taking into account additional information such as synonyms, word forms, and sentence structures~\cite{banerjee2005meteor}.\par

\renewcommand{\arraystretch}{1.7}
\begin{table*}[!t]
\centering
\caption{Performance of BugMentor - 550 samples}
\label{tab:bugmentorcontext2}
\begin{tabular}{|c|c|c|c|c|c|c|}
\hline
\textbf{Metrics} & \textbf{Top K} & \textbf{Python} & \textbf{Java} & \textbf{JavaScript} & \textbf{C++} & \textbf{WholeDataset} \\ \hline \hline
\multirow{3}{*}{\textbf{BLEU ↑ (0 - 100)}} & Top   - 1 &\textbf{34.38} & 31.28 & 28.3 & 27.97 & 30.48 \\ \cline{2-7} 
 & Top - 3 &\textbf{34.99} & 33.22 & 29.12 & 29.84 & 31.79 \\ \cline{2-7} 
 & Top - 5 &\textbf{35.25} & 36.18 & 31.12 & 30.05 & 33.15 \\ \hline
\multirow{3}{*}{\textbf{METEOR ↑ (0 - 1)}} & Top   - 1 & 0.68 &\textbf{0.79} & 0.66 & 0.62 & 0.69 \\ \cline{2-7} 
 & Top - 3 & 0.72  &\textbf{0.8} & 0.71 & 0.65 & 0.72 \\ \cline{2-7} 
 & Top - 5 & 0.76 &\textbf{0.83} & 0.73 & 0.69 & 0.75 \\ \hline
\multirow{3}{*}{\textbf{\begin{tabular}[c]{@{}c@{}}SEMANTIC\\      SIMILARITY ↑ (0 - 100)\end{tabular}}} & Top   - 1 & 63.01 &\textbf{79.94} & 54.71 & 50.49 & 62.04 \\ \cline{2-7} 
 & Top - 3 & 64.37 &\textbf{80.25} & 57.82 & 54.49 & 64.23 \\ \cline{2-7} 
 & Top - 5 & 70.11 &\textbf{82.35} & 59.69 & 57.06 & 67.3 \\ \hline
\multirow{3}{*}{\textbf{ROUGE ↑ (0 - 1)}} & Top   - 1 &\textbf{0.48} & 0.47 & 0.44 & 0.43 & 0.46 \\ \cline{2-7} 
 & Top - 3 &\textbf{0.53} & 0.51 & 0.47 & 0.44 & 0.49 \\ \cline{2-7} 
 & Top - 5 &\textbf{0.53} & 0.53 & 0.49 & 0.46 & 0.50 \\ \hline
\end{tabular}%
\end{table*}

As shown in Table~\ref{tab:bugmentorcontext2}, BugMentor achieves an average METEOR score of 0.69 for Top 1 answer, and a maximum of 0.75 for Top 5 answers against the whole dataset, which are considered to be reasonable~\cite{agarwal2008meteor}. Interestingly, it achieves a maximum of 0.79 for Top 1, 0.80 for Top 3, and 0.83 for Top 5 answers across four subsets. They show that BugMentor was able to produce a significant portion of the ground truth texts in the generated answers. However, since BLEU and METEOR scores rely on keyword matching between a generated answer and the ground truth answer, they may not capture the semantic relevance between them. Hence, we also evaluate our answers using ROUGE~\cite{huang2016supervised} and Semantic Similarity~\cite{haque2022semantic}. They have also been shown to correlate better with human judgement of relevance ~\cite{papineni2002bleu, kusner2015word}. \par

\renewcommand{\arraystretch}{1.7}
\begin{table*}[!ht]
\centering
\caption{Performance of BugMentor - 1887 samples}
\label{tab:bugmentor1887}
\begin{tabular}{|c|c|c|c|c|c|c|}
\hline
\textbf{Metrics} & \textbf{Top K} & \textbf{Python} & \textbf{Java} & \textbf{JavaScript} & \textbf{C++} & \textbf{WholeDataset} \\ \hline \hline
\multirow{3}{*}{\textbf{BLEU ↑}} & Top - 1 & \textbf{66.50} & 26.40 & 27.84 & 41.78 & 40.63 \\ \cline{2-7} 
 & Top - 3 & \textbf{72.08} & 26.40 & 28.27 & 41.78 & 42.13 \\ \cline{2-7} 
 & Top - 5 & \textbf{72.08} & 31.73 & 28.27 & 45.14 & 44.31 \\ \hline
\multirow{3}{*}{\textbf{METEOR ↑}} & Top - 1 & \textbf{0.77} & 0.47 & 0.40 & 0.74 & 0.60 \\ \cline{2-7} 
 & Top - 3 & \textbf{0.81} & 0.47 & 0.43 & 0.75 & 0.61 \\ \cline{2-7} 
 & Top - 5 & \textbf{0.81} & 0.52 & 0.43 & 0.75 & 0.63 \\ \hline
\multirow{3}{*}{\textbf{\begin{tabular}[c]{@{}c@{}}SEMANTIC\\ SIMILARITY ↑\end{tabular}}} & Top - 1 & \textbf{91.00} & 40.39 & 49.27 & 77.50 & 64.54 \\ \cline{2-7} 
 & Top - 3 & \textbf{92.50} & 44.23 & 50.13 & 78.07 & 66.23 \\ \cline{2-7} 
 & Top - 5 & \textbf{92.50} & 44.23 & 50.13 & 78.07 & 66.23 \\ \hline
\multirow{3}{*}{\textbf{ROUGE ↑}} & Top - 1 & \textbf{0.72} & 0.43 & 0.32 & 0.67 & 0.54 \\ \cline{2-7} 
\multicolumn{1}{|l|}{} & Top - 3 & \textbf{0.77} & 0.43 & 0.32 & 0.67 & 0.55 \\ \cline{2-7} 
\multicolumn{1}{|l|}{} & Top - 5 & \textbf{0.77} & 0.47 & 0.32 & 0.68 & 0.56 \\ \hline
\end{tabular}%
\end{table*}
\looseness=-1
In Table~\ref{tab:bugmentorcontext2}, we find that BugMentor achieves an average ROUGE of 0.46 for Top 1 answer, and a maximum of 0.50 for Top 5 answers, when the whole dataset is considered. It achieves a maximum of 0.48 for Top 1, 0.53 for Top 3, and 0.53 for Top 5 answers across four subsets. These scores show that BugMentor was able to generate answers semantically similar to the ground truth and were worded closely to the ground truth. However, ROUGE has limitations in capturing semantic nuances and contextual importance of words ~\cite{lin2004rouge,ng2015better}. It primarily measures surface-level text similarity and may miss valid paraphrases or alternative expressions that humans would recognize as equivalent. We thus also evaluate our answers using semantic similarity against the ground truth. The metric is appropriate when there may not be any syntactic overlap between the answers, which is a common phenomenon in question answering, according to existing literature ~\cite{yao2013answer,severyn2015learning}.\par

\looseness=-1
From Table ~\ref{tab:bugmentorcontext2}, we also find that the answers from BugMentor have an average Semantic Similarity score of  62.04\% for Top 1 answer, and a maximum of 67.30\% for Top 5 answers when the whole dataset is considered. BugMentor achieves a maximum of 79.94\% for Top 1, and a maximum of 82.35\% for Top 5 answers when all subsets are considered. All these numbers indicate a high similarity in meaning and content between BugMentor's answers and the ground truth.\par

BugMentor's performance varies across programming languages. It consistently shows the best results across most metrics for the Java subset, particularly excelling in METEOR and Semantic Similarity. For the Python subset, BugMentor follows a close second, achieving high BLEU scores and demonstrating consistent strength in other metrics. However, BugMentor's performance for the JavaScript and C++ subsets is generally lower compared to its results for Java and Python subsets. This could be due to several factors. JavaScript's dynamic and loosely-typed nature may lead to more complex and ambiguous bug descriptions, making it harder to generate accurate answers~\cite{jensen2009type}. C++'s low-level control and manual memory management may result in more intricate bugs that are difficult to address with our generated answers~\cite{nilsen1995real,stroustrup1986overview}. 

Across all languages, BugMentor's performance improves when considering Top-3 and Top-5 results compared to Top-1, suggesting increased accuracy with more potential solutions. These patterns may be influenced by factors such as the complexity of the bugs, or the structure of the languages themselves. Languages like Java and Python, which have large, active communities, verbose and expressive syntax, and established ecosystems, may lead to more comprehensive and well-structured bug reports, resulting in better performance for BugMentor. On the other hand, languages with more dynamic and flexible features, like JavaScript~\cite{jensen2009type}, or those with complex low-level control, like C++~\cite{nilsen1995real,stroustrup1986overview}, may present challenges in generating accurate answers due to the potential ambiguity and complexity in bug descriptions. Considering these additional factors provides a more better understanding of their potential influences on BugMentor's performance across different programming languages.

We also extend our dataset from  550 samples (95\% confidence level, 4.06\% margin of error) to 1887 samples (98\% confidence level, 2.5\% margin of error). We repeat our experiment with the extended dataset and report the results. The improved sampling parameters and extended samples provide greater statistical power and better representation of diverse bug scenarios. Table~\ref{tab:bugmentor1887} contains our results for the extended dataset.

We find performance improvements when 1887 samples were considered. BugMentor achieves an average BLEU Score of 40.63 for Top 1 answer, and a maximum of 44.31 for Top 5 answers when our whole dataset is considered. These BLEU scores indicate that our generated answers are of high quality according to Google’s AutoML Translation documentation. BugMentor shows stronger results particularly in Python and C++, with Python achieving BLEU scores of 66.50 (Top-1) and C++ reaching 41.78 (Top-1), compared to their previous scores of 34.38 and 27.97 respectively. BugMentor's performance varies across programming languages. It consistently shows the best results across most metrics for the Python subset, excelling in BLEU and Semantic Similarity. For instance, Python achieves a Semantic Similarity score of 91.00 for Top-1 answer, significantly higher than other languages. 

BugMentor's performance varies across programming languages when 1887 samples were considered. We find the best results across all metrics for the Python systems. For the C++ subset, it also achieves high Semantic Similarity and METEOR scores while demonstrating consistent strength in other metrics. However, BugMentor's performance measures for the Java and JavaScript subsets are consistently lower. This could happen due to several factors. First, JavaScript's dynamic, loosely-typed nature may lead to complex bugs and consequently ambiguous bug descriptions, making it harder for BugMentor to generate appropriate answers~\cite{jensen2009type}. On the other hand,  Java's verbose syntax, strong type system, and object-oriented paradigm could lead to bugs spanning multiple program components. This could also make it harder for BugMentor to provide helpful answers to follow-up questions, especially when those questions involve complex class interactions or rely on deeper knowledge of the broader Java ecosystem.

We also conducted a cross-project evaluation to assess the generalizability of BugMentor, especially in scenarios where a new project may have limited or no historical bug reports available for retrieval. In our experiment, a target bug report was drawn from one project in a given programming language, and the retrieval corpus (i.e., the candidate pool) was composed of bug reports from other projects in the same programming language. We repeat this for every project in each of the four programming languages and evaluate BugMentor’s performance. We also compared it with three baselines --- ChatGPT, Lucene, and AnswerBot --- with the same setting.\par

Our results from these cross-project experiments are shown in Table~\ref{tab:bugmentor-cross}. We see that BugMentor consistently outperforms all baselines across languages and metrics, despite no access to in-project historical data.  However, the performance varies across the programming languages. For example, in Python, BugMentor delivers more accurate answers than the baselines, achieving a Top-1 BLEU score of 58.44, which is higher than those of ChatGPT (11.42), Lucene (5.38) and AnswerBot (6.93). We observe similar trends in the other programming languages. For C++, BugMentor achieves a Top-1 BLEU score of 34.44 compared to 7.35 of ChatGPT, and its best Semantic Similarity of 70.28 (at Top-3) compared to ChatGPT of 33.52 (at Top-5). Although BugMentor achieves lower cross-project scores in Java and JavaScript compared to Python, it still consistently outperforms the baselines in those languages. For example, in Java, BugMentor’s Top-5 BLEU is 27.74 compared to 8.38 of ChatGPT, and in JavaScript. The Semantic Similarity of BugMentor peaks at 44.95 (Top-3) compared to 29.11 of ChatGPT. These findings indicate that BugMentor is capable of working with projects having limited or no bug reporting history, making it feasible for adoption in new or low-resource software systems.\par

\begin{table*}[!ht]
\centering
\caption{Performance of BugMentor in Cross-Project Setting}\resizebox{\textwidth}{!}{%
\label{tab:bugmentor-cross}\begin{tabular}{|c|c|cccc|cccc|cccc|cccc|}
\hline
\multirow{2}{*}{\textbf{Metrics}} & \multicolumn{1}{l|}{\multirow{2}{*}{\textbf{Top K}}} & \multicolumn{4}{c|}{\textbf{Lucene}} & \multicolumn{4}{c|}{\textbf{AnswerBot}} & \multicolumn{4}{c|}{\textbf{ChatGPT}} & \multicolumn{4}{c|}{\textbf{BugMentor}} \\ \cline{3-18} 
 & \multicolumn{1}{l|}{} & \multicolumn{1}{c|}{\textbf{Python}} & \multicolumn{1}{c|}{\textbf{Java}} & \multicolumn{1}{c|}{\textbf{JavaScript}} & \textbf{C++} & \multicolumn{1}{c|}{\textbf{Python}} & \multicolumn{1}{c|}{\textbf{Java}} & \multicolumn{1}{c|}{\textbf{JavaScript}} & \textbf{C++} & \multicolumn{1}{c|}{\textbf{Python}} & \multicolumn{1}{c|}{\textbf{Java}} & \multicolumn{1}{c|}{\textbf{JavaScript}} & \textbf{C++} & \multicolumn{1}{c|}{\textbf{Python}} & \multicolumn{1}{c|}{\textbf{Java}} & \multicolumn{1}{c|}{\textbf{JavaScript}} & \textbf{C++} \\ \hline \hline
\multirow{3}{*}{\textbf{BLEU ↑}} & Top - 1 & \multicolumn{1}{c|}{5.38} & \multicolumn{1}{c|}{6.07} & \multicolumn{1}{c|}{0.70} & 7.20 & \multicolumn{1}{c|}{6.93} & \multicolumn{1}{c|}{6.77} & \multicolumn{1}{c|}{5.40} & 2.83 & \multicolumn{1}{c|}{11.42} & \multicolumn{1}{c|}{7.70} & \multicolumn{1}{c|}{13.09} & 7.35 & \multicolumn{1}{c|}{58.44} & \multicolumn{1}{c|}{22.50} & \multicolumn{1}{c|}{24.23} & 34.44 \\ \cline{2-18} 
 & Top - 3 & \multicolumn{1}{c|}{5.38} & \multicolumn{1}{c|}{6.07} & \multicolumn{1}{c|}{0.70} & 7.20 & \multicolumn{1}{c|}{7.96} & \multicolumn{1}{c|}{7.10} & \multicolumn{1}{c|}{5.40} & 6.10 & \multicolumn{1}{c|}{11.42} & \multicolumn{1}{c|}{8.02} & \multicolumn{1}{c|}{13.09} & 7.53 & \multicolumn{1}{c|}{62.70} & \multicolumn{1}{c|}{25.50} & \multicolumn{1}{c|}{26.94} & 36.37 \\ \cline{2-18} 
 & Top - 5 & \multicolumn{1}{c|}{6.46} & \multicolumn{1}{c|}{6.07} & \multicolumn{1}{c|}{0.70} & 7.20 & \multicolumn{1}{c|}{7.96} & \multicolumn{1}{c|}{7.10} & \multicolumn{1}{c|}{6.34} & 8.89 & \multicolumn{1}{c|}{13.7} & \multicolumn{1}{c|}{8.38} & \multicolumn{1}{c|}{15.71} & 7.78 & \multicolumn{1}{c|}{63.88} & \multicolumn{1}{c|}{27.74} & \multicolumn{1}{c|}{26.18} & 37.35 \\ \hline
\multirow{3}{*}{\textbf{METEOR ↑}} & Top - 1 & \multicolumn{1}{c|}{0.20} & \multicolumn{1}{c|}{0.06} & \multicolumn{1}{c|}{0.14} & 0.13 & \multicolumn{1}{c|}{0.27} & \multicolumn{1}{c|}{0.24} & \multicolumn{1}{c|}{0.21} & 0.26 & \multicolumn{1}{c|}{0.36} & \multicolumn{1}{c|}{0.32} & \multicolumn{1}{c|}{0.35} & 0.30 & \multicolumn{1}{c|}{0.68} & \multicolumn{1}{c|}{0.41} & \multicolumn{1}{c|}{0.34} & 0.60 \\ \cline{2-18} 
 & Top - 3 & \multicolumn{1}{c|}{0.21} & \multicolumn{1}{c|}{0.06} & \multicolumn{1}{c|}{0.14} & 0.13 & \multicolumn{1}{c|}{0.28} & \multicolumn{1}{c|}{0.24} & \multicolumn{1}{c|}{0.22} & 0.27 & \multicolumn{1}{c|}{0.36} & \multicolumn{1}{c|}{0.32} & \multicolumn{1}{c|}{0.35} & 0.30 & \multicolumn{1}{c|}{0.70} & \multicolumn{1}{c|}{0.43} & \multicolumn{1}{c|}{0.36} & 0.62 \\ \cline{2-18} 
 & Top - 5 & \multicolumn{1}{c|}{0.24} & \multicolumn{1}{c|}{0.06} & \multicolumn{1}{c|}{0.14} & 0.13 & \multicolumn{1}{c|}{0.29} & \multicolumn{1}{c|}{0.25} & \multicolumn{1}{c|}{0.23} & 0.27 & \multicolumn{1}{c|}{0.39} & \multicolumn{1}{c|}{0.32} & \multicolumn{1}{c|}{0.37} & 0.31 & \multicolumn{1}{c|}{0.71} & \multicolumn{1}{c|}{0.45} & \multicolumn{1}{c|}{0.36} & 0.63 \\ \hline
\multirow{3}{*}{\textbf{\begin{tabular}[c]{@{}c@{}}SEMANTIC\\ SIMILARITY ↑\end{tabular}}} & Top - 1 & \multicolumn{1}{c|}{28.97} & \multicolumn{1}{c|}{23.33} & \multicolumn{1}{c|}{25.76} & 25.79 & \multicolumn{1}{c|}{21.33} & \multicolumn{1}{c|}{24.31} & \multicolumn{1}{c|}{21.31} & 23.01 & \multicolumn{1}{c|}{40.32} & \multicolumn{1}{c|}{38.17} & \multicolumn{1}{c|}{27.47} & 32.61 & \multicolumn{1}{c|}{81.92} & \multicolumn{1}{c|}{38.82} & \multicolumn{1}{c|}{42.46} & 68.03 \\ \cline{2-18} 
 & Top - 3 & \multicolumn{1}{c|}{29.77} & \multicolumn{1}{c|}{23.75} & \multicolumn{1}{c|}{26.51} & 26.66 & \multicolumn{1}{c|}{23.47} & \multicolumn{1}{c|}{24.45} & \multicolumn{1}{c|}{21.34} & 23.58 & \multicolumn{1}{c|}{48.61} & \multicolumn{1}{c|}{38.27} & \multicolumn{1}{c|}{27.78} & 32.73 & \multicolumn{1}{c|}{81.92} & \multicolumn{1}{c|}{41.05} & \multicolumn{1}{c|}{44.95} & 70.28 \\ \cline{2-18} 
 & Top - 5 & \multicolumn{1}{c|}{34.05} & \multicolumn{1}{c|}{25.48} & \multicolumn{1}{c|}{30.44} & 31.61 & \multicolumn{1}{c|}{25.65} & \multicolumn{1}{c|}{24.45} & \multicolumn{1}{c|}{22.45} & 23.89 & \multicolumn{1}{c|}{45.98} & \multicolumn{1}{c|}{39.17} & \multicolumn{1}{c|}{29.11} & 33.52 & \multicolumn{1}{c|}{82.58} & \multicolumn{1}{c|}{41.49} & \multicolumn{1}{c|}{44.58} & 70.28 \\ \hline
\multirow{3}{*}{\textbf{ROUGE ↑}} & Top - 1 & \multicolumn{1}{c|}{0.36} & \multicolumn{1}{c|}{0.03} & \multicolumn{1}{c|}{0.24} & 0.25 & \multicolumn{1}{c|}{0.31} & \multicolumn{1}{c|}{0.26} & \multicolumn{1}{c|}{0.22} & 0.28 & \multicolumn{1}{c|}{0.37} & \multicolumn{1}{c|}{0.34} & \multicolumn{1}{c|}{0.35} & 0.33 & \multicolumn{1}{c|}{0.61} & \multicolumn{1}{c|}{0.39} & \multicolumn{1}{c|}{0.30} & 0.55 \\ \cline{2-18} 
 & Top - 3 & \multicolumn{1}{c|}{0.36} & \multicolumn{1}{c|}{0.03} & \multicolumn{1}{c|}{0.24} & 0.25 & \multicolumn{1}{c|}{0.31} & \multicolumn{1}{c|}{0.26} & \multicolumn{1}{c|}{0.23} & 0.29 & \multicolumn{1}{c|}{0.37} & \multicolumn{1}{c|}{0.35} & \multicolumn{1}{c|}{0.35} & 0.33 & \multicolumn{1}{c|}{0.65} & \multicolumn{1}{c|}{0.41} & \multicolumn{1}{c|}{0.32} & 0.56 \\ \cline{2-18} 
 & Top - 5 & \multicolumn{1}{c|}{0.42} & \multicolumn{1}{c|}{0.03} & \multicolumn{1}{c|}{0.24} & 0.25 & \multicolumn{1}{c|}{0.33} & \multicolumn{1}{c|}{0.27} & \multicolumn{1}{c|}{0.25} & 0.31 & \multicolumn{1}{c|}{0.40} & \multicolumn{1}{c|}{0.35} & \multicolumn{1}{c|}{0.37} & 0.34 & \multicolumn{1}{c|}{0.66} & \multicolumn{1}{c|}{0.42} & \multicolumn{1}{c|}{0.32} & 0.57 \\ \hline
\end{tabular}
}
\end{table*}

\begin{boxH}
\looseness=-1
\textbf{Summary of RQ$\mathbf{_1}$:} 
BugMentor achieves an average BLEU score of 44, which shows that it can generate relevant answers to follow-up questions that are \textit{high quality} according to Google's Standard. It achieves an average semantic similarity score of 67\%, which shows that the answers also have a high semantic overlap with the ground truth answers. We also find that BugMentor performs the best for Python across most metrics while suffering from low performance with projects written in Java, JavaScript and C++.
\end{boxH}

\subsection{Answering RQ$\mathbf{_2}$ --- Can BugMentor outperform the existing baselines in terms of automatic evaluation metrics?}\label{results:rq2}

\begin{table*}[!ht]
\centering
\caption{Comparison of BugMentor with Baseline$_{Lucene}$,  DialBERT, AnswerBot and Baseline$_{Mistral}$ - 550 samples}
\label{tab:baselinecomparison1}
\begin{tabular}{|c|c|c|c|c|c|c|}
\hline
\textbf{Metrics} & \textbf{Top K} & \textbf{Baseline$_{Lucene}$} & \textbf{DialBERT} & \textbf{AnswerBot} & \textbf{Baseline$_{Mistral}$} & \textbf{BugMentor} \\ \hline \hline
\multirow{3}{*}{\textbf{BLEU ↑}} & Top   - 1 & 18.02 & 9.85 & 4.31 & 20.69 & 30.48 \\ \cline{2-7} 
 & Top - 3 & 20.18 & 12.55 & 6.71 & \multirow{2}{*}{} & 31.79 \\ \cline{2-5} \cline{7-7} 
 & Top - 5 & 23.04 & 12.78 & 6.71 &  & 33.15 \\ \hline
\multirow{3}{*}{\textbf{METEOR ↑}} & Top   - 1 & 0.59 & 0.31 & 0.05 & 0.59 & 0.69 \\ \cline{2-7} 
 & Top - 3 & 0.63 & 0.34 & 0.12 & \multirow{2}{*}{} & 0.72 \\ \cline{2-5} \cline{7-7} 
 & Top - 5 & 0.64 & 0.38 & 0.12 &  & 0.75 \\ \hline
\multirow{3}{*}{\textbf{\begin{tabular}[c]{@{}c@{}}SEMANTIC\\      SIMILARITY ↑\end{tabular}}} & Top   - 1 & 48.18 & 61.38 & 36.47 & 57.02 & 62.04 \\ \cline{2-7} 
 & Top - 3 & 50.88 & 64.16 & 47.69 & \multirow{2}{*}{} & 64.23 \\ \cline{2-5} \cline{7-7} 
 & Top - 5 & 52.92 & 64.88 & 47.69 &  & 67.30 \\ \hline
\multirow{3}{*}{\textbf{ROUGE ↑}} & Top   - 1 & 0.43 & 0.43 & 0.16 & 0.42 & 0.46 \\ \cline{2-7} 
 & Top - 3 & 0.45 & 0.46 & 0.17 & \multirow{2}{*}{} & 0.49 \\ \cline{2-5} \cline{7-7} 
 & Top - 5 & 0.45 & 0.46 & 0.19 &  & 0.50 \\ \hline
\end{tabular}%
\end{table*}

\begin{table*}[!ht]
\centering
\caption{Comparison of BugMentor with Baseline$_{Lucene}$,  DialBERT, AnswerBot and Baseline$_{Mistral}$ - 1887 samples}
\label{tab:baselinecomparison2}
\begin{tabular}{|c|c|c|c|c|l|c|}
\hline
\textbf{Metrics} & \textbf{Top K} & \textbf{Baseline\_Lucene} & \textbf{DialBERT} & \textbf{AnswerBot} & \multicolumn{1}{c|}{\textbf{Baseline\_Mistral}} & \textbf{BugMentor} \\ \hline \hline
\multirow{3}{*}{\textbf{BLEU ↑}} & Top - 1 & 21.25 & 18.40 & 14.40 & \multicolumn{1}{c|}{27.15} & \textbf{40.63} \\ \cline{2-7} 
 & Top - 3 & 23.34 & 20.80 & 15.66 & \multirow{2}{*}{} & \textbf{42.13} \\ \cline{2-5} \cline{7-7} 
 & Top - 5 & 24.85 & 22.10 & 16.21 &  & \textbf{44.31} \\ \hline
\multirow{3}{*}{\textbf{METEOR ↑}} & Top - 1 & 0.41 & 0.45 & 0.19 & \multicolumn{1}{c|}{0.55} & \textbf{0.60} \\ \cline{2-7} 
 & Top - 3 & 0.43 & 0.48 & 0.19 & \multirow{2}{*}{} & \textbf{0.61} \\ \cline{2-5} \cline{7-7} 
 & Top - 5 & 0.45 & 0.50 & 0.21 &  & \textbf{0.63} \\ \hline
\multirow{3}{*}{\textbf{\begin{tabular}[c]{@{}c@{}}SEMANTIC\\ SIMILARITY ↑\end{tabular}}} & Top - 1 & 49.50 & 59.80 & 23.38 & \multicolumn{1}{c|}{41.90} & \textbf{64.54} \\ \cline{2-7} 
 & Top - 3 & 51.75 & 63.10 & 25.46 & \multirow{2}{*}{} & \textbf{66.23} \\ \cline{2-5} \cline{7-7} 
 & Top - 5 & 53.00 & 64.25 & 26.05 &  & \textbf{66.23} \\ \hline
\multirow{3}{*}{\textbf{ROUGE ↑}} & Top - 1 & 0.44 & 0.47 & 0.18 & \multicolumn{1}{c|}{0.51} & \textbf{0.54} \\ \cline{2-7} 
 & Top - 3 & 0.46 & 0.49 & 0.19 & \multirow{2}{*}{} & \textbf{0.55} \\ \cline{2-5} \cline{7-7} 
 & Top - 5 & 0.47 & 0.51 & 0.20 &  & \textbf{0.56} \\ \hline
\end{tabular}%
\end{table*}
In this experiment, we evaluate the answers from all four baselines -- Lucene~\cite{mccandless2010lucene}, Mistral~\cite{jiang2023mistral}, AnswerBot~\cite{xu2017answerbot}, and DialBERT~\cite{li2020dialbert} -- against the ground truth using four evaluation metrics. Table~\ref{tab:baselinecomparison1} and Table~\ref{tab:baselinecomparison2} show the comparison details between BugMentor and these baseline techniques on 550 and 1887 samples, respectively. \par

BugMentor outperforms Baseline$_{Lucene}$ across all four metrics on the 550 dataset. BugMentor achieves 43\%-69\% performance improvement in terms of BLEU, 14\%-17\% in terms of METEOR, 26\%-28\% in terms of Semantic Similarity and 6\%-11\% in terms of ROUGE. Baseline$_{Mistral}$ also performs poorly compared to both BugMentor. BugMentor achieves a performance improvement of 47\% in terms of BLEU, 16\% in terms of METEOR, 8\% in terms of Semantic Similarity and 9\% in terms of ROUGE. When compared with DialBERT, BugMentor achieves 153\%-209\% performance improvement for BLEU, 97\%-122\% in terms of METEOR, 1\%-3\% in terms of Semantic Similarity and 6\%-8\% in terms of ROUGE. Similarly, Answerbot performs poorly when compared to BugMentor. BugMentor acheives a performance improvement of 373\%-607\% in terms of BLEU, 500\%-1200\% in terms of METEOR, 34\%-70\% in terms of Semantic Similarity and 163\%-
187\% in terms of ROUGE.\par

BugMentor also outperforms all baseline techniques across all four metrics, even with the extended dataset. Baseline$_{Mistral}$ and Baseline$_{Lucene}$ are the second closest in performance. Our proposed technique -- BugMentor -- outperforms Baseline$_{Mistral}$ by 49.65\% higher  BLEU, 9.09\% higher  METEOR,  54.03\% higher Semantic Similarity and 5.88\% higher ROGUE. On the other hand,  the performance improvements over Baseline$_{Lucene}$  are 78\%-91\% in terms of BLEU, 40\%-46\% in terms of METEOR, 24\%-30\% in terms of Semantic Similarity and 19\%-22\% in terms of ROUGE. When compared with DialBERT, BugMentor achieves 100\%-120\% performance improvement for BLEU, 26\%-33\% in terms of METEOR, 3\%-7\% in terms of Semantic Similarity and 9\%-14\% in terms of ROUGE.  Similarly, BugMentor outperforms AnswerBot by a large margin.





\looseness=-1
We also perform Mann-Whitney Wilcoxon test~\cite{cuzick1985wilcoxon} to check if the performance of Baseline$_{Lucene}$, Baseline$_{Mistral}$, DialBERT and AnswerBot are significantly lower than that of BugMentor using BonFerroni Correction~\cite{weisstein2004bonferroni} on both 550 and 1887 datasets. We find that BugMentor performs significantly higher than Baseline$_{Lucene}$, AnswerBot and DialBERT, i.e., p-value $=$ 0.00016, 00016 and 0.0016 $<$0.016 in terms of all four metrics for 550 dataset and p-value = 0.00016, 0.0016 and 0.00016 <0.016 in terms of all four metrics for the 1887 dataset. To quantify the magnitude of the differences, we calculate Cliff's Delta effect sizes~\cite{cohen1988statistical}. The effect sizes for BugMentor compared to Baseline${Lucene}$, AnswerBot, and DialBERT are 0.25, 0.38 and 0.32, for the 550 dataset and 0.38, 0.50 and 0.38 for the 1887 dataset, indicating small to medium effect.\par


\begin{boxH}
\looseness=-1
\textbf{Summary of RQ$\mathbf{_2}$:} 
BugMentor performs better in answer generation than all four baselines in terms of four evaluation metrics. According to the statistical significance test, BugMentor outperforms the closest competitors Baseline$_{Lucene}$, DialBERT and AnswerBot with statistical significance and small to medium effect.
\end{boxH}

\subsection{Answering RQ$\mathbf{_3}$ --- How do different components impact the overall performance of BugMentor?}

\looseness=-1
To answers a follow-up question our technique combines three key components --- (a) structured information retrieval, (b) embedding similarity-based ranking and (c) neural text generation. 
In this experiment, we conduct an ablation study to determine the contribution of each component. In particular, we design different variants of BugMentor with each component and evaluate their performance in answer generation. Table~\ref{tab:ablation} summarizes our results from the ablation study.\par

\looseness=-1
We find that BugMentor outperforms all of its variants based on either individual components or their combinations. The combination of structured information retrieval and embedding similarity-based ranking (a.k.a., BugMentor$_{L+E}$) is a close second when compared using BLEU and ROUGE scores.\par

\renewcommand{\arraystretch}{1.6}
\begin{table}[!t]
\begin{center}
\caption{Experimental Results from the Ablation Study}
\label{tab:ablation}
\resizebox{\textwidth}{!}{%
\begin{tabular}{|c|c|c|c|c|l|c|c|c|c|}
\hline
\textbf{Metrics} & \textbf{Top K} & \textbf{BugMentor$_{L}$} & \textbf{BugMentor$_{L+E}$} & \textbf{BugMentor$_{L+M}$} & \multicolumn{1}{c|}{\textbf{BugMentor$_{M}$}} & \textbf{BugMentor$_{E}$} & \textbf{BugMentor$_{E+M}$} & \textbf{BugMentor$_{RO+M}$} & \textbf{BugMentor} \\ \hline \hline
\multirow{3}{*}{\textbf{BLEU ↑}} & Top - 1 & 21.25 & 24.93 & 8.68 & \multicolumn{1}{c|}{20.69} & 21.1 & 25.2 & 10.98 & 30.48 \\ \cline{2-10} 
 & Top - 3 & 23.47 & 26.63 & 9.33 & \multirow{2}{*}{} & 23.3 & 26.9 & 10.98 & 31.79 \\ \cline{2-5} \cline{7-10} 
 & Top - 5 & 24.48 & 29.35 & 9.79 &  & 24.7 & 28.3 & 11.18 & 33.15 \\ \hline
\multirow{3}{*}{\textbf{METEOR ↑}} & Top - 1 & 0.31 & 0.52 & 0.41 & \multicolumn{1}{c|}{0.59} & 0.46 & 0.58 & 0.35 & 0.69 \\ \cline{2-10} 
 & Top - 3 & 0.33 & 0.55 & 0.43 & \multirow{2}{*}{} & 0.48 & 0.61 & 0.35 & 0.72 \\ \cline{2-5} \cline{7-10} 
 & Top - 5 & 0.35 & 0.57 & 0.44 &  & 0.5 & 0.63 & 0.36 & 0.75 \\ \hline
\multirow{3}{*}{\textbf{\begin{tabular}[c]{@{}c@{}}SEMANTIC\\ SIMILARITY ↑\end{tabular}}} & Top - 1 & 48.5 & 55.39 & 33.56 & \multicolumn{1}{c|}{73.61} & 55.2 & 61.9 & 27.71 & 62.04 \\ \cline{2-10} 
 & Top - 3 & 50.98 & 56.3 & 35.15 & \multirow{2}{*}{} & 57.5 & 62.6 & 27.84 & 64.23 \\ \cline{2-5} \cline{7-10} 
 & Top - 5 & 53.51 & 57.02 & 35.84 &  & 59.8 & 65.01 & 27.89 & 67.3 \\ \hline
\multirow{3}{*}{\textbf{ROUGE ↑}} & Top - 1 & 0.41 & 0.44 & 0.34 & \multicolumn{1}{c|}{0.42} & 0.4 & 0.44 & 0.35 & 0.46 \\ \cline{2-10} 
 & Top - 3 & 0.45 & 0.47 & 0.36 & \multirow{2}{*}{} & 0.43 & 0.46 & 0.35 & 0.49 \\ \cline{2-5} \cline{7-10} 
 & Top - 5 & 0.46 & 0.49 & 0.36 &  & 0.44 & 0.47 & 0.35 & 0.5 \\ \hline
\end{tabular}
}
\end{center}
\begin{tablenotes}
  \item \begin{center}
     \textbf{BugMentor$_{L}$} $=$ BugMentor$_{Lucene}$, \textbf{BugMentor$_{E}$} $=$ BugMentor$_{Embedding}$, \textbf{BugMentor$_{M}$} $=$ BugMentor$_{Mistral}$, 
     \textbf{BugMentor$_{L+E}$} $=$ BugMentor$_{Lucene + Embedding}$ , \textbf{BugMentor$_{L+M}$} $=$ BugMentor$_{Lucene + Mistral}$,
     \textbf{BugMentor$_{E+M}$} $=$ BugMentor$_{Embedding + Mistral}$ ,
     \textbf{BugMentor$_{RO+M}$} $=$ BugMentor$_{Retrieval Only + Mistral}$
  \end{center}
\end{tablenotes}
\end{table}

\looseness=-1
From Table~\ref{tab:ablation}, we find that the combination of Embedding and Mistral is the closest competitor to BugMentor. However, the absence of the text generation component (a.k.a., Mistral) and structured information retrieval component (a.k.a Lucene) reduces the performance from that of BugMentor by 44.45\%  in terms of BLEU, 50\% in terms of METEOR,  12.39\% in terms of Semantic Similarity and 15\% in terms of ROUGE. We also note that the combination of Lucene and Embedding (a.k.a., BugMentor$_{L + E}$) is the second closest competitor to BugMentor when BLEU is considered. However, the absence of the text generation component (a.k.a., Mistral) reduces the performance from that of BugMentor by 18.20\% in terms of BLEU and 10.71\% in terms of Semantic Similarity when the Top 1 answer is captured and 4.08\% in terms of ROUGE when Top 3 is captured. In other words, the text generation component improves the semantic closeness between the generated answers and the ground truth answer. Thus, the Mistral component has a significant impact on improving not only the understandability of our generated answers (i.e., BLEU) but also their syntactic and semantic relevance to the ground truth answers (i.e. METEOR, ROUGE, Semantic Similarity).\par

\looseness=-1
We also include a retrieval-only prompting baseline, BugMentor$_{RO+M}$, which directly supplies the top-ranked historical Q\&A pair to the LLM without any structured reasoning or context integration. Theoretically, this lightweight strategy serves as a lower bound for our approach. As expected, this baseline (i.e., BugMentor$_{RO+M}$) performs significantly worse than the full BugMentor. For instance, BugMentor$_{RO+M}$ achieves a lower BLEU of 10.98 for the Top-1 result compared to 30.48 for BugMentor. Similarly, inferior METEOR and Semantic Similarity metrics were observed for the baseline (Table~\ref{tab:ablation}). This pronounced gap in performance confirms that simply prompting the LLM with a single retrieved Q\&A pair is insufficient, underscoring the importance of BugMentor’s combination of structured retrieval, ranking, and context-enhanced generation for robust performance.\par

\looseness=-1
From Table~\ref{tab:ablation}, we also find that the components from BugMentor do not perform well when evaluated individually. For example, if we omit both Embedding and Mistral components and only use Lucene, the performance degrades by 30.28\% in terms of BLEU, 55.07\% in terms of METEOR, 21.82\% in terms of Semantic Similarity and 10.86\% in terms of ROUGE for the Top 1 answer. Similarly, if we use only the Mistral component, the performance degrades by 32.11\% in terms of BLEU, 14.49\% in terms of METEOR and 8.09\% in terms of Semantic Similarity and 8.69\% in terms of ROUGE for Top 1 answer. Thus, these components in isolation might not be sufficient to deliver appropriate answers to the follow-up questions from bug reports.\par

We also investigated BugMentor’s compatibility with different backbone LLMs. In particular, we replaced the Mistral component with the Llama variant and repeated our experiments. Table~\ref{tab:ablation2} summarises our results.

\begin{table*}[!ht]
\centering
\caption{Comparison of BugMentor with Baseline$_{LLama}$, BugMentor$_{LLama}$ and Baseline$_{Mistral}$ }
\label{tab:ablation2}
\begin{tabular}{|c|c|c|c|c|c|}
\hline
\textbf{Metrics} & \textbf{Top K} & \textbf{BugMentor$_{LLama}$} & \textbf{BugMentor} & \textbf{Baseline$_{LLama}$} & \textbf{Baseline$_{Mistral}$} \\ \hline \hline
\multirow{3}{*}{\textbf{BLEU ↑}} & Top - 1 & 52.8 & 40.63 & 30.38 & 27.15 \\ \cline{2-6} 
 & Top - 3 & 54.35 & 42.13 & \multirow{2}{*}{} & \multirow{2}{*}{} \\ \cline{2-4}
 & Top - 5 & 56.1 & 44.31 &  &  \\ \hline
\multirow{3}{*}{\textbf{METEOR ↑}} & Top - 1 & 0.64 & 0.6 & 0.49 & 0.55 \\ \cline{2-6} 
 & Top - 3 & 0.66 & 0.61 & \multirow{2}{*}{} & \multirow{2}{*}{} \\ \cline{2-4}
 & Top - 5 & 0.68 & 0.63 &  &  \\ \hline
\multirow{3}{*}{\begin{tabular}[c]{@{}c@{}}\textbf{SEMANTIC}\\ \textbf{SIMILARITY ↑}\end{tabular}} & Top - 1 & 71.1 & 64.54 & 67.91 & 61.9 \\ \cline{2-6} 
 & Top - 3 & 72.5 & 66.23 & \multirow{2}{*}{} & \multirow{2}{*}{} \\ \cline{2-4}
 & Top - 5 & 72.5 & 66.23 &  &  \\ \hline
\multirow{3}{*}{\textbf{ROUGE ↑}} & Top - 1 & 0.58 & 0.54 & 0.48 & 0.51 \\ \cline{2-6} 
 & Top - 3 & 0.58 & 0.55 & \multirow{2}{*}{} & \multirow{2}{*}{} \\ \cline{2-4}
 & Top - 5 & 0.62 & 0.56 &  &  \\ \hline
\end{tabular}
\end{table*}
Although we initially selected Mistral for the neural text generation component due to its practical advantages (e.g., lower memory consumption and faster inference compared to Llama~\cite{mistral2023blog_mistral7b}), we extended our experiments to compare with Llama. Specifically, we evaluated Baseline${Mistral}$, Baseline${Llama}$, BugMentor${Llama}$, and Baseline${Mistral}$ on the extended dataset. As shown in Table 1, both Baseline models perform poorly in isolation, but the addition of BugMentor’s retrieval and ranking context yields substantial improvements. BugMentor improves BLEU scores by 49.65\% over standalone Mistral, while BugMentor${Llama}$ improves by 33.73\% over standalone Llama. Importantly, BugMentor${Llama}$ achieves higher absolute performance across BLEU, METEOR, and Semantic Similarity than BugMentor${Mistral}$, highlighting Llama’s strong potential for bug-related answer generation.\par

Beyond accuracy, deployment cost was an important consideration for us. Mistral’s lower memory footprint and faster inference make it well-suited for resource-constrained environments (e.g., edge devices or lightweight IDEs). In contrast, Llama, being larger by design, requires greater computational resources, yet consistently delivers stronger empirical performance across all metrics. These findings highlight that BugMentor is compatible with both models — Llama offers superior accuracy when resources permit, while Mistral provides a cost-effective option for efficiency-critical deployments.\par

\begin{boxH}
\looseness=-1
\textbf{Summary of RQ$\mathbf{_3}$:} 
Our ablation study demonstrates significant contribution of the three components --- structured information retrieval, embedding similarity-based ranking, and neural text generation (a.k.a., LLM backbone) --- towards the overall performance of BugMentor. We also found that BugMentor outperforms its five variants based on individual components, their combinations, or even a retrieval-only baseline, which justifies the presence of all three components in BugMentor. We also found that BugMentor is model-agnostic and can deliver superior performance in question answering with different LLM backbones (e.g., Mistral, Llama).
\end{boxH}


\subsection{Answering RQ$\mathbf{_4}$:  How does BugMentor perform in answering follow-up questions when compared with ChatGPT?}\label{results:rq4}

Besides the traditional baselines, it is important to consider the emerging, popular LLM-based approaches such as ChatGPT for a comparative analysis. 
Since the most recent model of ChatGPT i.e., GPT-4o is a paid model, we conduct a quantitative analysis using ChatGPT. We use 550 bug reports $\sim$27 bug reports x 20 systems) for our analysis. First, we collect the bug reports, follow-up questions, ground truth answers and the answers generated by BugMentor. Then, we provide each follow-up question and its corresponding bug report in the same prompt and collect the answers from ChatGPT for our comparison. Table~\ref{tab:chatgpt} and Table~\ref{tab:chatgpt2} summarise our comparison details. Additionally, we replace the Mistral component of BugMentor with the ChatGPT component and repeat our experiments. We conduct experiments to determine the performance when past bug reports are provided by BugMentor for contextual information. 

\renewcommand{\arraystretch}{1.4}
\begin{table}[htbp]
\caption{Comparison between BugMentor$_{ChatGPT}$, BugMentor and ChatGPT - 550 samples}
\label{tab:chatgpt}
\centering
\begin{tabular}{|c|c|c|c|c|}
\hline
\textbf{Metrics} & \textbf{Top K} & \textbf{BugMentor$_{ChatGPT}$} & \textbf{BugMentor} & \multicolumn{1}{l|}{\textbf{ChatGPT}} \\ \hline \hline
\multirow{3}{*}{\textbf{BLEU ↑}} & Top - 1 & 52.35 & 30.48 & 20.09 \\ \cline{2-5} 
\multicolumn{1}{|l|}{} & Top - 3 & 54.72 & 31.79 & \multirow{2}{*}{} \\ \cline{2-4}
\multicolumn{1}{|l|}{} & Top - 5 & 56.89 & 33.15 &  \\ \hline
\multirow{3}{*}{\textbf{METEOR ↑}} & Top - 1 & 0.753 & 0.69 & 0.46 \\ \cline{2-5} 
 & Top - 3 & 0.781 & 0.72 & \multirow{2}{*}{} \\ \cline{2-4}
 & Top - 5 & 0.802 & 0.75 &  \\ \hline
\multirow{3}{*}{\textbf{\begin{tabular}[c]{@{}c@{}}SEMANTIC\\ SIMILARITY ↑\end{tabular}}} & Top - 1 & 82.45 & 62.04 & 65.64 \\ \cline{2-5} 
 & Top - 3 & 85.12 & 64.23 & \multirow{2}{*}{} \\ \cline{2-4}
 & Top - 5 & 88.37 & 67.3 &  \\ \hline
\multirow{3}{*}{\textbf{ROUGE ↑}} & Top - 1 & 0.502 & 0.46 & 0.48 \\ \cline{2-5} 
 & Top - 3 & 0.531 & 0.49 & \multicolumn{1}{l|}{\multirow{2}{*}{}} \\ \cline{2-4}
 & Top - 5 & 0.553 & 0.50 & \multicolumn{1}{l|}{} \\ \hline
\end{tabular}
\end{table}

\renewcommand{\arraystretch}{1.4}
\begin{table}[htbp]
\caption{Comparison between BugMentor$_{ChatGPT}$, BugMentor and ChatGPT - 1887 samples}
\label{tab:chatgpt2}
\centering
\begin{tabular}{|c|c|c|c|l|}
\hline
\textbf{Metrics} & \textbf{Top K} & \textbf{BugMentor$_{ChatGPT}$} & \textbf{BugMentor} & \multicolumn{1}{c|}{\textbf{ChatGPT}} \\ \hline \hline
\multirow{3}{*}{\textbf{BLEU ↑}} & Top - 1 & 36.03 & \textbf{40.63} & \multicolumn{1}{c|}{28.57} \\ \cline{2-5} 
 & Top - 3 & 38.59 & \textbf{42.13} & \multirow{2}{*}{} \\ \cline{2-4}
 & Top - 5 & \textbf{44.8} & 44.31 &  \\ \hline
\multirow{3}{*}{\textbf{METEOR ↑}} & Top - 1 & 0.46 & \textbf{0.6} & \multicolumn{1}{c|}{0.51} \\ \cline{2-5} 
 & Top - 3 & 0.49 & \textbf{0.61} & \multirow{2}{*}{} \\ \cline{2-4}
 & Top - 5 & 0.53 & \textbf{0.63} &  \\ \hline
\multirow{3}{*}{\textbf{\begin{tabular}[c]{@{}c@{}}SEMANTIC\\ SIMILARITY ↑\end{tabular}}} & Top - 1 & \textbf{69.79} & 64.54 & \multicolumn{1}{c|}{57.75} \\ \cline{2-5} 
 & Top - 3 & \textbf{71.74} & 66.23 & \multirow{2}{*}{} \\ \cline{2-4}
 & Top - 5 & \textbf{77.42} & 66.23 &  \\ \hline
\multirow{3}{*}{\textbf{ROUGE ↑}} & Top - 1 & \textbf{0.56} & 0.54 & \multicolumn{1}{c|}{0.54} \\ \cline{2-5} 
 & Top - 3 & \textbf{0.58} & 0.55 & \multirow{2}{*}{} \\ \cline{2-4}
 & Top - 5 & \textbf{0.61} & 0.56 &  \\ \hline
\end{tabular}
\end{table}

From Table ~\ref{tab:chatgpt}, we observe that ChatGPT is able to achieve a BLEU score of 13.22 which indicates that the answers from ChatGPT does not have a significant overlap with the ground truth text. We also find the METEOR score, since it takes into account synonyms, word forms and sentence structures. It achieves a METEOR score of 0.34 which shows that the answers have a poor keyword overlap. We also evaluate ChatGPT's answers using Semantic Similarity and ROUGE scores which better captures semantic and contextual importance of words. We observe that the answers from ChatGPT achieve a Semantic Similarity of 43.99 and ROUGE score of 0.41 which indicates a low semantic overlap with the ground truth answers. In contrast, BugMentor demonstrates a significant performance improvement across all metrics. Compared to ChatGPT, BugMentor achieves a 130.5\% improvement in terms of BLEU score, a 102.9\% improvement in METEOR score, a 41.03\% improvement in Semantic Similarity, and a 12\% improvement in ROUGE score. These results highlight BugMentor's superior ability to generate answers that closely resemble the ground truth, both in terms of syntactic and semantic similarity.\par

We compare the performance of three systems -- BugMentor$_{ChatGPT}$, BugMentor, and ChatGPT -- across four metrics for various Top-K answers for both 550 and 1887 datasets(Table~\ref {tab:chatgpt} and Table~\ref{tab:chatgpt2}). BugMentor$_{ChatGPT}$ consistently outperforms the other two systems across all metrics and Top-K answers for the 550 datasets. For example, in the BLEU metric, which measures text accuracy and fluency, BugMentor$_{ChatGPT}$ achieves a score of 52.35 at Top-1, significantly higher than BugMentor (30.48) and ChatGPT (20.09). This trend continues at Top-3 and Top-5, demonstrating the model’s ability to generate more accurate answers consistently. Similarly, in the METEOR metric, which evaluates text quality based on precision, recall, and alignment, BugMentor$_{ChatGPT}$ scores 0.802 at Top-5, compared to 0.75 for BugMentor, indicating better semantic and syntactic alignment with reference texts. This suggests that BugMentor$_{ChatGPT}$ produces responses that are not only more grammatically and syntactically accurate but also semantically richer and closer to the intended meaning.  In the ROUGE metric, which measures the overlap of n-grams between generated and reference texts, BugMentor$_{ChatGPT}$ also performs better, scoring 0.553 at Top-5 compared to 0.50 for BugMentor. This indicates that BugMentor$_{ChatGPT}$ generates more relevant and precise text. 

We compare the performance of three systems for the 1887 datasets. We find that BugMentor consistently outperforms the other two methods across BLEU and METEOR. For example, in terms of METEOR, BugMentor outperforms BugMentor$_{ChatGPT}$ by 18\%-30\% for Top 1,3 and 5 results. Similarly, in terms of BLEU, BugMentor outperforms BugMentor$_{ChatGPT}$ by 9-12\% for Top 1 and Top 3. This suggests that BugMentor generates answers that are syntactically accurate, semantically richer and closer to the ground truth texts. However, we find BugMentor$_{ChatGPT}$ performs marginally better in the Top 5 results of BLEU by 1\%. However, we observe that BugMentor$_{ChatGPT}$ outperforms BugMentor in terms of Semantic Similarity by 7-14\% and ROUGE by 3-8\%. Thus, BugMentor provides more accurate responses that align closely with reference solutions, while BugMentor$_{ChatGPT}$ covers a wider range of semantic variations. As shown above, our approach can work with various language models, indicating its flexibility and modularity.\par

To demonstrate the benefit of BugMentor, we present a use case (Bug Report \#1182) and compare with ChatGPT head to head below. The bug report discusses a performance issue with image detection in TorchServe, where inference times were significantly impacted by network delays when transferring large images. Despite being on the same network as TorchServe, the reporter experienced transfer times of 2-3 minutes before pre-processing could begin, compared to the actual inference time of 10-20 seconds. However, the bug report does not provide sufficient details. A developer asks more details about the setup and how the transfer time versus the pre-processing time are measured and if TorchServe is running on the same machine, where the images are located. The ground truth reveals that TorchServe runs on the same machine as the images, with data transferred over localhost. The reporter measures time by logging timestamps at various points, finding a 2-minute delay between sending the request and the start of pre-processing. The reporter also provides their machine specifications. From the below example we can see that BugMentor's Answer is more aligned with the ground truth, and suggests logging at different stages, capturing the context of the bug report. Addtionally, BugMentor's answer provides additional context about the performance differences between WSL and Docker that are not mentioned in the ground truth. In comparison, ChatGPT is also able to capture the context of transfer and pre-processing time and suggests logging, however, it does not provide additional information about the performance difference between Docker on Windows and WSL. \par

\begin{boxH}
\looseness=-1
\textbf{Summary of RQ$\mathbf{_4}$:} 
BugMentor outperforms ChatGPT in answering follow-up questions for a sample of 228 bug reports across four languages. While ChatGPT understands bug context, it lacks specificity in its answers. BugMentor, by learning from historical bug data, provides more accurate and contextual responses, surpassing ChatGPT on metrics like BLEU, METEOR, Semantic Similarity, and ROUGE scores.
\end{boxH}

\begin{shaded*}
\noindent
\begin{center}
    \textbf{Bug ~\hyperref[gpt-ans]{\#1182:}}
    \label{gpt-ans}
\end{center}
\textbf{Title:} 
Slow Connections with Image Detection \\

\noindent
\textbf{Description:}
Is your feature request related to a problem? Please describe:
Inference times with image detection seem to be slowed down by network delays with large images. I'm on the same network as TorchServe, but transferring large images to TorchServe takes about 2-3 minutes before preprocessing even starts, compared to 10-20 seconds for inference. I could move the preprocessing outside of TorchServe, but that would defeat the purpose of having all model-related code contained by TorchServe. Even then, my preprocessing only resizes the images. \\
Describe the solution:
Can the code include strategies or options to reduce network transfer times of images, possibly by changing the connection type?
Describe alternative solutions:
The documentation could also be updated to show how to mitigate this issue outside of TorchServe.\\

\noindent 
\textbf{Question:}
Interesting, I think this would be valuable for us to repro if you have anything we could use.
Is torchserve running on the same machine as where your images are? Any details you can provide would be super helpful
How are you measuring the transfer time vs preprocess time?\\

\noindent
\textbf{Ground Truth:}
Torchserve is running on the same machine as where my images are (and on the same disk!). So all data is being transferred over localhost connections.
In my torchserve handler, I'm printing the time as soon as it hits the`preprocess()`function. Then similar time prints allow me to measure the time spent on`preprocess()`,`inference()`, and`postprocess()`.

The main outcome here is that the time from just before I send the request to the`preprocess()`function starting tends to be about 2 minutes on average.

I'm using an efficientdet d4 object detection model, but I don't think the type of model should matter based on this information?

My machine specs are windows10 (running in WSL), Ryzen 7 3700x, 16GB RAM, GTX1060 6GB \\

\noindent
\textbf{Bugmentor Answer}
TorchServe is running on the same machine as the images, with data transferred over localhost. To measure the transfer delay, the time is logged before sending the image and when preprocessing starts. The observed delay is around 2 minutes. Notably, this issue doesn’t occur in WSL 2 but does in Windows, suggesting a performance difference between Docker on Windows and WSL. This aligns with a previous case where WSL showed notably better performance. \\

\noindent
\textbf{ChatGPT Answer}
TorchServe is not running on the same machine as the images; they're transferred over the network. To measure transfer time versus preprocessing time, you should log timestamps at the start and end of each stage: when the image transfer begins and ends, and when preprocessing starts and finishes. This will help pinpoint where delays are occurring.
\end{shaded*}

\subsection{Answering RQ$\mathbf{_5}$ --- How accurate, precise, useful, and concise are the answers from BugMentor?}\label{section:devstudy}

The metric-based evaluation above demonstrates the benefits of our technique, BugMentor,  in answering follow-up questions from bug reports. We conduct a preliminary developer study involving 23 participants to further demonstrate the benefits of BugMentor. Given a bug report (e.g., title, description) and its follow-up question, we evaluate how \emph{accurate}, \emph{precise}, \emph{useful}, and \emph{concise} (details in Table~\ref{tab:devstudyqualityaspects}) our answers are according to human developers. Although the developer study provided useful practitioner feedback, we acknowledge that its scope was limited. The study was conducted using only 12 bug reports, which represent a small sample compared to the diversity and complexity of real-world software bugs. Due to this limitation, our results should not be interpreted as broadly generalizable.\par

\begin{table}[h]
\centering
\caption{Quality Aspects of the Generated Answers}
\label{tab:devstudyqualityaspects}
\renewcommand{\arraystretch}{1.5}
\begin{tabular}{|l|l|}
\hline
\textbf{Quality} & \textbf{Overview} \\ \hline \hline
Accurate  & It provides the same factual information as the reference. \\ \hline
Precise & It can answer the question completely \\ \hline
Concise & It is short and still answers the question. \\ \hline
Useful & The provided information has the potential to answer the question. \\ \hline
\end{tabular}
\end{table}

\begin{table}[!t]
\small
\centering
\caption{Comparison of BugMentor with the Baseline Techniques using a Developer Study}
\label{tab:devstudy}
\setlength{\tabcolsep}{5pt}
\begin{tabular}{|c|c|c|c|c|}
\hline
\textbf{Quality} & \textbf{Model} & \textbf{Mean} & \textbf{Median} & \textbf{Mode} \\ \hline \hline
\multirow{4}{*}{Accurate} & Baseline$_{Lucene}$ & 2.63 & 3 & 3 \\ \cline{2-5} 
 & Baseline$_{Mistral}$ & 2.33 & 2 & 1 \\ \cline{2-5} 
 & ChatGPT & 2.38 & 2 & 1 \\ \cline{2-5} 
 & \textbf{BugMentor} & \textbf{3.01} & \textbf{3} & \textbf{3} \\ \hline
\multirow{4}{*}{Precise} & Baseline$_{Lucene}$ & 2.60 & 3 & 3 \\ \cline{2-5} 
 & Baseline$_{Mistral}$ & 2.36 & 2 & 2 \\ \cline{2-5} 
 & ChatGPT & 2.31 & 2 & 1 \\ \cline{2-5} 
 & \textbf{BugMentor} & \textbf{3.03} & \textbf{3} & \textbf{3} \\ \hline
\multirow{4}{*}{Useful} & Baseline$_{Lucene}$ & 2.53 & 3 & 3 \\ \cline{2-5} 
 & Baseline$_{Mistral}$ & 2.43 & 2 & 2 \\ \cline{2-5} 
 & ChatGPT & 2.42 & 2 & 2 \\ \cline{2-5} 
 & \textbf{BugMentor} & \textbf{3.71} & \textbf{4} & \textbf{5} \\ \hline
\multirow{4}{*}{Concise} & Baseline$_{Lucene}$ & 2.45 & 2 & 2 \\ \cline{2-5} 
 & Baseline$_{Mistral}$ & 2.35 & 2 & 2 \\ \cline{2-5} 
 & ChatGPT & 2.34 & 2 & 3 \\ \cline{2-5} 
 & \textbf{BugMentor} & \textbf{3.64} & \textbf{4} & \textbf{4} \\ \hline
\end{tabular}%
\end{table}

\textbf{Study Participants:} 
\looseness=-1
For our study, we target English-speaking software developers with professional experience in four programming languages --- Python, Java, JS, and C++ --- that were used by our collected projects. We sent an open invitation through our personal connections, and 10 graduate students and 13 working professionals responded. Each of them had professional experience in software development and at least two years of experience in the aforementioned programming languages. We provide them with a quick overview of our project using relevant examples of bug reports, follow-up questions and answers from GitHub. We provide them a secured link~\cite{OpinioSurvey} containing our survey on Opinio, a tool for conducting anonymous surveys. None of the participants knew the specifics of our designed technique --- BugMentor. \par

\looseness=-1
\textbf{Study Setup:} For our study, we use 12 use cases where each use case consists of a bug report and a follow-up question. To select these use cases, we apply random sampling without replacement to the held-out dataset (Section~\ref{sec:dataset-cons}). To avoid information overload, we select such bug reports that (a) do not have any stack trace information and (b) do not warrant any project-specific knowledge to understand the bug. We select three randomly sampled bug reports from each subset (based on programming language, Section~\ref{sec:dataset-cons}) and collect corresponding follow-up questions, ground truth answers, and the answers from both BugMentor and baseline techniques --- Baseline$_{Lucene}$,  Baseline$_{Mistral}$ and ChatGPT.\par

\looseness=-1
We present all 12 use cases to each of the participants. Then the participants were instructed to assess the accuracy, precision, usefulness and conciseness of the generated answers (by BugMentor and baselines) with respect to the ground truth answers. We instruct the participants to submit their evaluation on a five-point Likert scale, where 1 indicates strongly disagree and 5 indicates strongly agree. \textbf{We also anonymize the source of all generated answers to avoid any potential bias towards any of the techniques.}
We thus collect a total of  192 data points (12 questions × 4 explanations × 4 evaluation aspects) from each of the 23 participants.\par

The design of our developer survey has been reviewed and approved by the Dalhousie University Research Ethics Board (REB file \#: 2023-6885).\par

\begin{figure*}[!t]
\centering
 \includegraphics[width=\textwidth, height=7in, keepaspectratio]{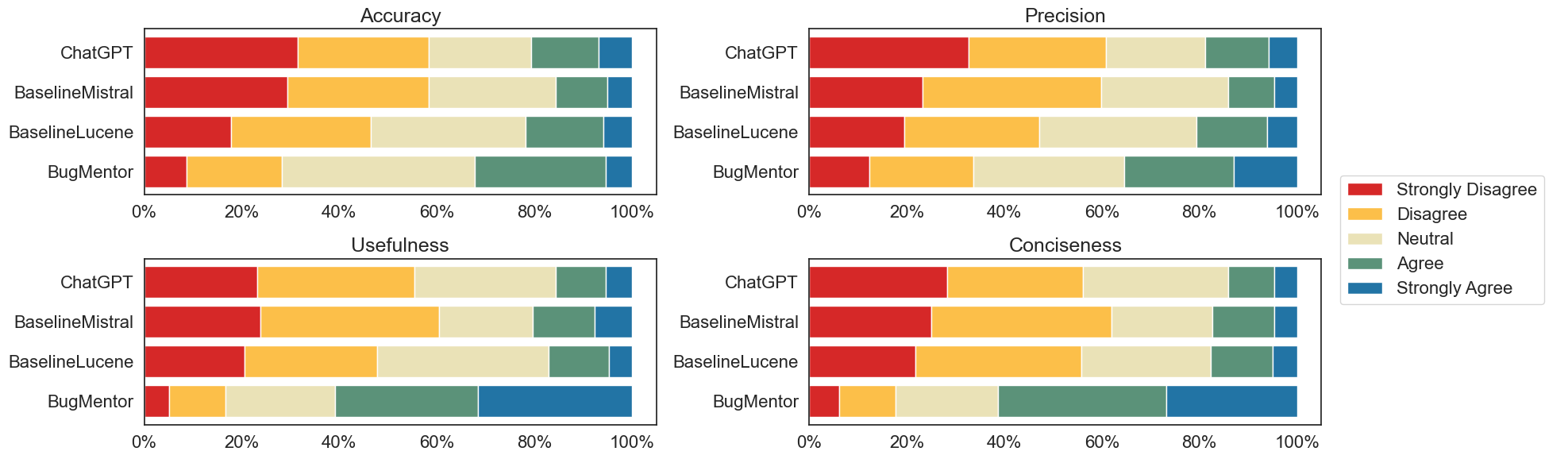}
	\caption{Comparison of BugMentor with the Baseline Techniques and ChatGPT using the Likert scores}
	\vspace{-.6cm}
	\label{fig:dev}
\end{figure*}

\textbf{Study Results and Discussion:} Table~\ref{tab:devstudy} summarizes our findings from the developer study. We note that, on average, the participants found the answers from BugMentor to be the most accurate, precise, useful, and concise. Based on the median and mode values, we see that the participants agree with our answers the most. Similar to the findings in RQ$\mathbf{_2}$, the participants found the closest competitor of BugMentor to be Baseline$_{Lucene}$ across all four metrics followed by ChatGPT. According to the median values, the developers agree with Baseline$_{Lucene}$ and BugMentor equally in terms of Accuracy and Precision.

\looseness=-1
Fig.~\ref{fig:dev} shows the distribution of participants’ agreement levels with different quality aspects of the answers. 
We observe that participants find BugMentor's answers to be most useful ($\sim$60\%) and most concise ($\sim$60\%) compared to Baseline$_{Lucene}$, Baseline$_{Mistral}$ and ChatGPT.
We also see that the participants strongly agree with BugMentor for a substantial part of the time (e.g., $\sim$30\% for accuracy), and strongly disagree only a few times (e.g., $<$20\% times), which none of the baselines achieved. On the other hand, nearly half of the time, the participants disagree with AnswerBot, Baseline$_{Lucene}$ and Baseline$_{Mistral}$ regarding various quality aspects such as Accuracy, Precision, Usefulness and Conciseness of their provided answers.\par

\looseness=-1
We also perform Mann-Whitney Wilcoxon test~\cite{cuzick1985wilcoxon} to check if the developers' preferences to Baseline$_{Lucene}$, Baseline$_{Mistral}$ and ChatGPT are significantly lower than that of BugMentor using BonFerroni Correction~\cite{weisstein2004bonferroni}. We find that the preference levels for BugMentor are significantly higher than all three baselines, i.e., p $=$ 0.0037$<$0.016 for Baseline$_{Lucene}$, p $=$ 0.0016$<$0.016 for  Baseline$_{Mistral}$, and p $=$ 0.00016$<$0.016 for ChatGPT. To quantify the magnitude of the differences, we calculate Cliff’s Delta Effect sizes for BugMentor compared to BaselineLucene, Mistral, and ChatGPT are 0.79, 1 and 0.97, indicating a large effect.

\textbf{Manual Analysis:}
\looseness=-1
To further investigate the usefulness of BugMentor's answers, we perform a manual analysis on 228 bug reports (i.e. 57 samples for each programming language) and 976 bug reports (i.e. 244 samples for each programming language). We select these samples from the whole dataset using stratified random sampling with a 95\% confidence level and 5\% error margin for 228 bug reports and 97.5\% confidence level and a 2.5\% margin of error for 976 samples. We collect the bug reports, follow-up questions, ground truth and generated answers.  For each bug report, we evaluate BugMentor's answers against the ground truth answers provided by the developer community, categorizing them into four levels of effectiveness -- 'Answers Completely' indicates responses that fully address the question with accurate technical details, 'Answers Partially' represents responses that contain correct but incomplete information; 'Additional Information' denotes cases where BugMentor provided relevant supplementary details beyond the ground truth, and 'Not Relevant' identifies responses that failed to address the core question or contained incorrect information. This categorization helped us systematically assess BugMentor's performance across different types of bug reports and programming languages. The primary manual analysis for the paper was conducted by the first author and documented using an Excel sheet, with a total of $\sim$20 hours spent on the analysis. Table~\ref{tab:manualanalysis} and Table~\ref{tab:manualanalysis2} show the summary of our analysis. \par

\renewcommand{\arraystretch}{1.5}
\begin{table}[htbp]
\caption{Manual Analysis - 228 samples}
\label{tab:manualanalysis}
\centering
\begin{threeparttable}
\begin{tabular}{|c|c|c|c|c|c|}
\hline
\textbf{Dataset} & \textbf{AC} & \textbf{AP} & \textbf{AP+AddInfo} & \textbf{AddInfo} & \textbf{NR} \\ \hline \hline
Java & 11 & 18 & 12 & 10 & 6 \\ \hline
Python & 13 & 21 & 9 & 9 & 5 \\ \hline
C++ & 9 & 18 & 13 & 13 & 4 \\ \hline
JavaScript & 8 & 20 & 11 & 11 & 7 \\ \hline
\textbf{Average\%} & 20.92\% & 39.29\% & 22.96\% & 21.94\% & 11.22\% \\ \hline
\end{tabular}   
\begin{tablenotes}
  \small
  \item \begin{center}
     \textbf{AC} $=$ Answers Completely, \textbf{AP} $=$ Answers Partially,\\ \textbf{AddInfo} $=$ Additional Information,\\ \textbf{NR} $=$ Not Relevant 
  \end{center}
\end{tablenotes}
\end{threeparttable}
\end{table}

\looseness=-1
We analyze our generated answers to the follow-up questions, contrast them against the ground truth, and determine whether they respond to the question completely, partially, or simply provide additional information or are not relevant. From Table~\ref{tab:manualanalysis}, we find that BugMentor, on average, was able to answer the questions completely for 20.92\% of the cases from each programming language. It was able to answer 39.29\% of the questions partially, while 22.96\% of the cases it was able to answer the questions partially while adding complementary information. Furthermore, in 21.94\% cases, our technique delivered such answers that did not match with the ground truth answers but were complementary or somewhat relevant to the questions. 11.22\% of the cases, BugMentor generated answers that were not relevant. From Table~\ref{tab:manualanalysis2}, we find that BugMentor, on average, was able to answer questions completely for 21.52\% of the cases from each programming language. It was able to answer 27.66\% of the questions partially and add complementary information to 22.95\% of the questions. Furthermore, according to our manual analysis, in 15.68\% of cases, our technique delivered such answers that did not match the ground truth answers but were complementary or somewhat relevant to the questions. For 12.19\% of the cases, BugMentor’s provided answers were not relevant.

While BugMentor demonstrates good performance in generating meaningful answers to follow-up questions, there are cases where it underperforms. Table~\ref{tab:underperforming-examples} provides two such cases. Here, we analyze two negative cases, identify the potential causes, and suggest strategies for improvement.\par

In bug \#2830, a Hive Task Lock issue, BugMentor struggled due to a vaguely worded follow-up question –"there is some exception info or how to reproduce this situation". The bug report also failed to include essential information such as exception logs, environment details, and reproduction steps. To handle such cases, existing work (e.g., Imran et al~\cite{imran2021automatically}) can be leveraged to capture complementary information such as exception logs, steps to reproduce, or relevant code snippets. For cases where direct user input may not be immediately available, incorporating external knowledge sources like Stack Overflow or technical documentation can supplement the bug reports and thus can help BugMentor generate more meaningful answers.\par

In Bug \#2071, a Kubernetes ClassCastException, BugMentor struggled to provide actionable insights due to the absence of critical information such as stack trace details in the bug report. To address such limitations, existing work (e.g., Chaparro et al.~\cite{chaparro2017detecting}) can be leveraged to improve the handling of incomplete or ambiguous bug reports. For instance, techniques for automatically extracting or inferring missing information, such as stack traces or environment details, could be applied. Furthermore, integrating external knowledge sources like documentation could help supplement the bug report with relevant context or solutions, even when user-provided details are insufficient.\par

\begin{boxH}
\looseness=-1
\textbf{Summary of RQ$\mathbf{_5}$:} Developers with professional development experience found BugMentor’s answers to be accurate, precise, concise, and useful in the 12 evaluated cases. Their preference levels for BugMentor were higher than those of the three baseline techniques by a statistically significant margin. However, we acknowledge that these results are based on a small set of bug reports and might not be generalizable to complex real-world bugs.
\end{boxH}


\renewcommand{\arraystretch}{1.5}
\begin{table}[htbp]
\caption{Manual Analysis - 976 samples}
\label{tab:manualanalysis2}
\centering
\begin{threeparttable}
\begin{tabular}{|c|c|c|c|c|c|}
\hline
\textbf{Dataset} & \textbf{AC} & \textbf{AP} & \textbf{AP+AddInfo} & \textbf{AddInfo} & \textbf{NR} \\ \hline \hline
Java & 61 & 91 & 34 & 32 & 26 \\ \hline
Python & 63 & 52 & 50 & 41 & 38 \\ \hline
JavaScript & 59 & 48 & 66 & 45 & 26 \\ \hline
C++ & 27 & 79 & 74 & 35 & 29 \\ \hline
\textbf{Average \%} & 21.52\% & 27.66\% & 22.95\% & 15.68\% & 12.19\% \\ \hline
\end{tabular}  
\begin{tablenotes}
  \small
  \item \begin{center}
     \textbf{AC} $=$ Answers Completely, \textbf{AP} $=$ Answers Partially,\\ \textbf{AddInfo} $=$ Additional Information,\\ \textbf{NR} $=$ Not Relevant 
  \end{center}
\end{tablenotes}
\end{threeparttable}
\end{table}

\section{Practical Implications}
The practical adoption of BugMentor can be integrated into existing software development workflows in various ways, which we discuss below.

\renewcommand{\arraystretch}{1.3}
\begin{table}[]
\caption{Underperforming Examples}
\label{tab:underperforming-examples}
\centering
\resizebox{0.9\columnwidth}{!}{
\fontsize{20}{19}\selectfont
\begin{tabular}{|l|l|l|}
\hline
\textbf{Bug ID} &
  \multicolumn{1}{c|}{\textbf{2071}} &
  \multicolumn{1}{c|}{\textbf{2830}} \\ \hline \hline
\textbf{Bug Report} &
  \begin{tabular}[c]{@{}l@{}}Title: ClassCastException: class io.fabric8.kubernetes.api.model.\\ Description: Secret cannot be cast to \\ class io.fabric8.kubernetes.api.model.GenericKubernetesResource\\ I have an operator where I set up watcher \\ using the customResource to be able to watch \\ any kind of resource, something like:\\ javaCustomResourceDefinition\\ Context ctx = createCRDContext(""v1"", ""Secret""); .....\end{tabular} &
  \begin{tabular}[c]{@{}l@{}}Title:Use eScheduler to Schedule the Hive Task\\ Description: \\ Stopping the workflow will lock the table. \\ You need to manually delete the lock record in Hive metadata. \\ Is there any way to avoid this problem? \\ Thank you.\end{tabular} \\ \hline
\textbf{\begin{tabular}[c]{@{}l@{}}Relevant Candidate Answer\end{tabular}} &
  \begin{tabular}[c]{@{}l@{}}I wouldn't agree 100\% on this being a bug.\\ \textgreater If you are dealing with a known type, \\ why do you want to handle this as a \\ Raw/Unstructured/CustomResource type ….\end{tabular} &
  \begin{tabular}[c]{@{}l@{}}@caishunfeng but i think re-run should keep\\  only one workflow instance, because semantically \\ speaking, re-run is indeed the same task. I think \\ we just need to determine the assignment logic of the \\ two field values of startTime and Duration…\end{tabular} \\ \hline
\textbf{Question} &
  \begin{tabular}[c]{@{}l@{}}Could you please add eprintstacktrace,\\ and check what's the exception this\\ is showing me, this openjdk bit \\ server vm warning\end{tabular} &
  \begin{tabular}[c]{@{}l@{}}Are there any exception info \\ or how to reproduce this \\ situation ?\end{tabular} \\ \hline
\textbf{Answer} &
  \begin{tabular}[c]{@{}l@{}}1. The bug report is not relevant to the question.\\ 2. The question is not clear.\\ 3. The question is not concise.\\ 4. The question is not relevant to the bug report.\\ 5. The question is not relevant to the relevant information.\\ 6. The question is not relevant to the relevant information.\\ 7. The question is not relevant to the relevant information.\end{tabular} &
  \begin{tabular}[c]{@{}l@{}}The exception is related to the use of the Hive task in the workflow. \\ When the Hive task is stopped, it will lock the table in the Hive\\ database. This causes the workflow to fail because the table is \\ locked. You need to manually delete the record of the lock \\ table in the Hive metadata. There is no way to avoid this \\ problem, but you can use...\end{tabular} \\ \hline
\end{tabular}}
\end{table}

\textbf{Addressing the Bottleneck in Bug Resolution}
BugMentor addresses a critical bottleneck in software maintenance: the timely resolution of follow-up questions in bug reports. Missing information in bug reports is a pervasive issue, with 77\% of developers from major tech companies identifying it as a major problem~\cite{zou2018practitioners}. By automating the generation of relevant answers to follow-up questions, BugMentor reduces the time developers spend waiting for responses and improves the efficiency of bug resolution. This directly tackles the challenge of non-reproducibility and incomplete bug reports, which are key factors behind delayed bug fixes~\cite{rahman2020some}.\par

\textbf{Enhancing Developer Productivity}
Given that developers spend approximately 50\% of their programming time on bug-related tasks~\cite{britton2013reversible}, BugMentor significantly enhances productivity. It provides immediate answers to follow-up questions, reduces back-and-forth communications, and prevents premature bug closures due to missing information. This not only streamlines the bug resolution process but also allows developers to focus on higher-value tasks, ultimately improving software quality and reducing maintenance costs.\par

\textbf{Seamless Integration into Existing Workflows}
BugMentor can be seamlessly integrated into existing software development workflows through multiple channels. It can be implemented as a plugin or extension for popular issue-tracking systems like GitHub Issues and JIRA. In these platforms, it automatically generates and suggests relevant answers when developers post follow-up questions on bug reports. This integration ensures that BugMentor becomes a natural part of the development process without requiring significant changes to existing workflows.\par

\textbf{Deployment as a Microservice}
For organizations with custom development infrastructures, BugMentor can be deployed as a microservice accessible through REST APIs. This makes it compatible with a wide range of development tools and environments, ensuring flexibility and scalability. The system's reliance on structured information retrieval and neural text generation (e.g., Mistral) ensures that it can handle diverse programming languages and project sizes, as demonstrated by its evaluation from top GitHub projects.\par

\textbf{Cost-effective and Scalable Adoption}
BugMentor's design ensures cost-effective adoption across projects of any size. It requires minimal additional infrastructure beyond existing bug-tracking systems, making it accessible to both small teams and large organizations. The provided replication package, which includes the working prototype, experimental dataset, and configuration details, further simplifies deployment and customization. This fosters broader adoption and enables organizations to tailor BugMentor to their specific needs.\par

\textbf{Reducing the Economic Impact of Software Bugs}
By automating the answering of follow-up questions, BugMentor contributes to cost-effective software maintenance. This aligns with the broader goal of reducing the economic impact of software bugs, which cost the global economy billions of dollars annually~\cite{zou2018practitioners,britton2013reversible}. The tool’s ability to generate accurate, precise, and concise answers, as validated by developer studies, ensures that it delivers tangible value in real-world scenarios. \par

\textbf{Facilitating Community and Industry Adoption}
The comprehensive evaluation of BugMentor, including its performance metrics (e.g., BLEU score, METEOR, ROUGE, Semantic Similarity) and developer studies, demonstrates its effectiveness across diverse programming languages and project sizes. This evidence-based validation, combined with the availability of a replication package, encourages adoption not only within individual organizations but also across the broader software development community. By addressing a universal pain point in software maintenance, BugMentor has the potential to become a widely adopted tool in both industry and open-source projects.\par

\section{Related Work}

\looseness=-1
Question Answering (QA) has been an active research topic in both Information Retrieval (IR) and Natural Language Processing (NLP) communities~\cite{ravichandran2002learning, brill2002analysis,waltz1978english, iyyer2014neural,asaduzzaman2013answering, tian2017apibot, lu2021beat, bansal2021neural, xu2017answerbot, abdellatif2020msrbot}. There have also been several works that focus on question-answering in the context of software engineering. Breu et al.~\cite{breu2010information} first analyzed follow-up questions from bug reports and found that 32.34\% of them were never responded to, leading to delays in fixing the bug. Their study highlighted the need for techniques to address the problem of follow-up questions and improve the bug resolution workflow. Recently, Imran et al.~\cite{imran2021automatically} proposed Bug-AutoQ that recommends follow-up questions against a deficient bug report leveraging development history using information retrieval. Their technique identifies relevant follow-up questions based on the similarity between the current bug report and the most similar historical bug report. Their goal was to assist developers in gathering relevant information. However, their technique does not answer the follow-up questions.\par

\looseness=-1
Murgia et al.~\cite{murgia2016among} conducted a preliminary study that leveraged the search feature of the StackOverflow Q\&A site to suggest relevant questions in response to error messages from version control systems like GitHub. While their technique showed promise, it was limited to providing only simple, recurring questions related to Git error messages, overlooking other types of errors or bugs. Tian et al.~\cite{tian2017apibot} proposed APIBot, a system capable of answering questions related to specific APIs by analyzing the relevant API documentation. APIBot demonstrated the potential of leveraging existing documentation to provide targeted answers. However, the scope of their solution was limited to API-related questions, overlooking other problems that developers encounter during the software development process. Bansal et al.~\cite{bansal2021neural} design a context-aware QA system to answer basic questions about subroutines. They designed an encoder-decoder model that extracts information directly from the source code. Lu et al.~\cite{lu2021beat} introduced a novel question-answering approach called BEAT (Bug rEport AssistanT) that aims to provide accurate and relevant answers to questions related to bug reports. Their approach leverages the structure and content of bug report templates to generate structured queries from complex questions that can be executed to retrieve relevant information effectively. However, their approach might fail when a bug report does not contain the requested information. Xu et al.~\cite{xu2017answerbot} designed AnswerBot to synthesize answers for technical, non-factoid questions related to Java from StackOverflow using information retrieval. However, they only use the title of a question overlooking the detailed problem context (e.g., question body), and thus their answers might be unaware of the problem context. We compare BugMentor with AnswerBot using experiments, and the detailed comparison can be found in Section~\ref{results:rq2}. Abdellatif et al.~\cite{abdellatif2020msrbot} introduced MSRBot, a system designed to answer the most common questions related to software development and maintenance. While MSRBot demonstrated the potential of mining software repositories to provide relevant answers, its effectiveness might be limited by the available information in the mined repositories. Song et al.~\cite{song2022toward} developed BURT, a system specifically designed to support bug reporters of Android applications. While BURT showed promise in assisting Android developers, its approach might not generalize well to other software applications, limiting its broader applicability.

Recently, Language Model-based approaches (LLMs) such as ChatGPT have emerged as a powerful text generation tool. After conducting a limited qualitative analysis (Section~\ref{results:rq2}), we note that while ChatGPT exhibit an understanding of a given bug report, they often struggle to come up with precise answers to follow-up questions. BugMentor has a better understanding of past bug reports and thus captures a broader context. We also do a thorough qualitative and quantitative analysis to compare the performances of BugMentor and ChatGPT. We find that while ChatGPT understands the bug context, it lacks specificity and additional information that BugMentor provides. The detailed comparison can be found in Section~\ref{results:rq4}

\looseness=-1
In summary, existing research in the field of automated question answering for software development has primarily focused on two main areas: improving deficient bug reports and addressing specific types of questions. Studies have proposed techniques to enhance the quality of bug reports by suggesting relevant questions or providing answers to inquiries related to APIs, subroutines, and Git error messages. While these contributions have been valuable, they have not directly tackled the challenge of automatically answering follow-up questions that arise from bug reports. To the best of our knowledge, our proposed technique is the first to automatically answer the follow-up questions from bug reports, which makes our work \emph{novel}. We also combine structured information retrieval with neural text generation (e.g., Mistral) to generate the answers, which were found to be meaningful, accurate, precise, useful, and concise according to two types of evaluation --- automated metrics and developer study. Our technique also outperforms four baselines from the existing literature, which demonstrates its strong potential.

\looseness=-1
We also distinguish our work from duplicate bug report detection, which focuses on identifying similar bug reports to avoid redundant maintenance efforts~\cite{nguyen2012duplicate,chaparro2019reformulating}. Our work differs from duplicate bug detection in several aspects. First, although duplicates are identified in our work, we leverage them to generate answers against follow-up questions from deficient bug reports, which is a novel research problem. According to existing work~\cite{rahman2020some},  unanswered questions pose significant maintenance challenges and lead to delayed bug resolution. Second, duplicate detection methods often rely on textual similarity or clustering techniques to identify the reports describing the same issue. In contrast, our work leverages contextual information and historical data from past, similar bug reports and employs neural text generation to synthesize meaningful responses to follow-up questions. Developers also found our answers to be useful, as shown by the developer study in Section~\ref{section:devstudy}. Finally, while duplicate bug detection aims to reduce redundancy, our work enhances the quality and completeness of bug reports by offering actionable answers to follow-up questions. Thus, our work attempts to improve bug report management by ensuring that the developers have the necessary information to triage and debug the bugs effectively.

\section{Threats to Validity}
We identify a few threats to the validity of our findings. In this section, we examine these threats and discuss the steps that were taken to mitigate them.\par

\looseness=-1
\textbf{External Validity:}
Threats to external validity refer to the lack of generalizability in the findings~\cite{ferguson2004external}. One threat could stem from our selection of subject systems. We select 20 software systems written in four programming languages: Python, Java, JavaScript, and C++, which might not represent all systems at GitHub. However, the underlying algorithm of BugMentor is not bound to any programming language and thus can be easily adapted to any other platforms.\par
The second threat could stem from the small sample size of the held-out dataset for evaluation (e.g., 550). However, to mitigate this concern, we selected them carefully through random sampling from all four subsets (95\% confidence level, 4.06\% error margin -- Section~\ref{sec:groundtruth}). We also maintain diversity in selecting our 20 subject systems (Section~\ref{sec:groundtruth}). Another threat could stem from the manual analysis; the constrained sample size may limit the generalizability of the findings. However, to mitigate this concern, we carefully random sample them from all four subsets using a 97.5\% confidence level and 2.5\% error margin.

\looseness=-1
\textbf{Construct Validity:} Construct validity refers to the extent to which the experiment measures what it intends to measure~\cite{smith2005construct}. 
Inappropriate use of evaluation metrics could be a threat to construct validity. However, we chose our evaluation metrics --- BLEU, METEOR, Semantic Similarity, and ROUGE --- based on relevant literature~\cite{papineni2002bleu,banerjee2005meteor,haque2022semantic,lin2004rouge}. We also chose the four quality aspects of generated answers based on relevant literature~\cite{imran2021automatically,joshi2015likert}. Thus, threats to construct validity might be mitigated.

\looseness=-1
\textbf{Internal Validity:} Threats to internal validity relate to experimental errors and subjective biases~\cite{christ2007experimental}. We use manually annotated ground truth to answer both RQ1 and RQ2, which could be a source of threat. However, to mitigate this, the annotators were given appropriate training for their annotation tasks. We also employ majority voting~\cite{kuhrmann2017pragmatic} for decision-making and calculate Cohen's $\kappa$ to demonstrate the agreement levels between annotators~\cite{kuhrmann2017pragmatic}.
In the developer study, the assessment of answers can be influenced by subjective bias. However, we anonymise the source of all answers to avoid any bias towards any technique. For our supplementary manual analysis, while it provided insights into BugMentor's answer quality, two limitations should be noted. The evaluation relied on a single rater, which may introduce subjective bias, and sample size constraints could limit generalizability.

\looseness=-1
Another source of threat could be the replication of the baseline techniques. For the replication of Mistral, we collected the pre-trained model from HuggingFace~\cite{huggingface}, and for the replication of Lucene, we used ElasticSearch~\cite{ElasticSearch}, a standard library. To replicate AnswerBot~\cite{xu2017answerbot}, we used the replication package from the original authors~\cite{maxxbw54}. To replicate DialBERT~\cite{li2020dialbert}, we used the replication package from the original author. Furthermore, we followed the documentation closely for any customizations. Thus, threats to internal validity might be mitigated.

\section{Conclusions and Future work}
Missing information in bug reports makes bug resolution challenging. Hence, developers often ask follow-up questions to gather additional information. This paper proposes BugMentor, an approach that answers the follow-up questions from bug reports combining structured information retrieval and neural text generation. Our technique leverages the relevance between past and current bug reports to gather additional context, which helps us generate an appropriate answer to the question. We evaluate the answers from BugMentor using both evaluation metrics and developer study. Our technique achieves an average BLEU Score of up to 36 and a Semantic Similarity of up to 82, demonstrating its ability to generate understandable and good answers to follow-up questions, as per Google's Standard. We further conduct a developer survey involving 23 participants to evaluate the answers. The participants found the answers from BugMentor to be more accurate, precise, concise and useful compared to the baseline answers.\par
\looseness=-1
There are several avenues for future work. First, we plan to design a tool encapsulating BugMentor and integrate into real-world platforms like GitHub or JIRA to assist the bug reporters and the developers in their work. In particular, feedback from real-life problem contexts will be useful to improve our algorithm.
Second, BugMentor does not use fine-tuned version of Mistral. If the labelled dataset can be extended, it can be used to fine-tune our designed model, which could lead to better answers.

\section*{Data Availability Statement (DAS)}

All the data generated or analyzed during this study are available in the
GitHub repository to help reproduce our results~\cite{replicationpackage}.

\bibliographystyle{plainnat}
\bibliography{refs}
\end{document}